\documentclass[aps,prx,showpacs,floatfix,twocolumn,superscriptaddress,nofootinbib,longbibliography]{revtex4-2}
\usepackage{graphicx}
\usepackage{bm,amsmath}
\usepackage{dcolumn}
\usepackage{amsfonts,amssymb}
\usepackage{mhchem}
\usepackage{diagbox}
\usepackage{subfigure}
\usepackage{bbm}
\usepackage[mathlines]{lineno}
\usepackage{hyperref}
\begin{document}
\title{Dynamical localization transition  in the  non-Hermitian lattice gauge theory}

\author{Jun-Qing Cheng}
\affiliation{State Key Laboratory of Optoelectronic Materials and Technologies, Guangdong Provincial Key Laboratory of Magnetoelectric Physics and Devices, School of Physics, Sun Yat-Sen University, Guangzhou 510275, China}
\affiliation{School of Physical Sciences, Great Bay University, Dongguan 523000, China, and Great Bay Institute for Advanced Study, Dongguan 523000, China}

\author{Shuai Yin}
\email{yinsh6@mail.sysu.edu.cn}
\affiliation{State Key Laboratory of Optoelectronic Materials and Technologies, Guangdong Provincial Key Laboratory of Magnetoelectric Physics and Devices, School of Physics, Sun Yat-Sen University, Guangzhou 510275, China}
\date{\today}

\author{Dao-Xin Yao}
\email{yaodaox@mail.sysu.edu.cn}
\affiliation{State Key Laboratory of Optoelectronic Materials and Technologies, Guangdong Provincial Key Laboratory of Magnetoelectric Physics and Devices, School of Physics, Sun Yat-Sen University, Guangzhou 510275, China}
\begin{abstract}

\section{abstract}
Local constraint in the lattice gauge theory provides an exotic mechanism that facilitates the disorder-free localization.  However, the understanding of nonequilibrium dynamics in the non-Hermitian lattice gauge model remains limited. Here, we investigate the quench dynamics in a system of spinless fermions with nonreciprocal hopping in the $\mathbb{Z}_2$ gauge field. By employing a duality mapping, we systematically explore the non-Hermitian skin effect, localization-delocalization transition, and real-complex transition. Through the identification of diverse scaling behaviors of quantum mutual information for fermions and spins, we propose that the non-Hermitian quantum disentangled liquids exist both in the localized and delocalized phases, the former originates from the $\mathbb{Z}_2$ gauge field and the latter arises from the non-Hermitian skin effect. Furthermore, we demonstrate that the nonreciprocal dissipation causes the flow of quantum information. Our results provide valuable insights into the nonequilibrium dynamics in the gauge field, and may be experimentally validated using quantum simulators.

\end{abstract}

\maketitle

\section{ Introduction}
Gauge  theories play a pivotal  role in providing a unified description of standard model in particle physics.  In recent decades, gauge theories have also found applicability in condensed matter physics \cite{wen2004quantum,Fradkin2013FieldTO}, particularly in the context of the many-body systems, leading to the emergence of lattice gauge theories \cite{PhysRevD.10.2445}. Prominent examples include  the lattice gauge descriptions of quantum magnets \citep{PhysRevLett.62.1694}, high-Tc superconductors \cite{PhysRevB.37.580} and quantum simulation \cite{martinez2016real,PhysRevLett.118.070501,schweizer2019,NP2019realization,science.JianweiPan,pan_2020,Banuls2020,science367,scienceadvance2019,PhysRevResearch.4.L022060,PhysRevX.10.021041,PhysRevResearch.4.033120}.  The local gauge symmetry inherent in these theories gives rise to a multitude of conserved quantities, making them a valuable platform for investigating the localization phenomena in the absence of disorder \citep{PhysRevLett.118.266601,thesis,PhysRevLett.120.030601,PhysRevB.102.104302,PhysRevA.103.022416,PRXQuantum.3.020345,halimeh2022temperature}. The lattice gauge theory framework holds great promise for the development of specialized digital and analogue quantum simulators, with the ultimate goal of realizing universal quantum computers. Consequently, this field has garnered significant attention in experimental research. Currently, the realization of ${\mathbb{Z}}_{2}$ lattice gauge theory has  been successfully achieved using diverse quantum resources such as  ultracold atoms \cite{schweizer2019,scienceadvance2019}, superconducting circuits \cite{PhysRevResearch.4.L022060}, and Rydberg atomic arrays \cite{PhysRevX.10.021041}. These experiments hold profound significance in guiding  the study of disorder-free  localization. 

Recently, the non-Hermitian physics has attached widespread attention due to the discovery of numerous  physical properties that transcend those exhibited by the Hermitian systems \cite{RevModPhys.88.035002,ElGanainy2018,Mirieaar7709}. For instance, the non-Hermitian  skin effect  has unveiled the  sensitivity to the spatial
boundary conditions \cite{PhysRevLett.116.133903,PhysRevLett.121.086803,PhysRevLett.121.026808}, the non-Hermiticity has expanded the realm of topological phases beyond the Hermitian framework \cite{PhysRevLett.121.086803,PhysRevLett.121.026808,PhysRevX.9.041015}, the interplay of non-Hermiticity and disorder  has given rise to the emergence of  the localization-delocalization transitions \cite{PhysRevLett.123.090603,PhysRevB.100.054301,PhysRevB.102.064206,PhysRevLett.126.166801}.  Non-Hermitian physics finds wide-ranging applicability across various  physical systems, including optics \cite{Naturephysics2010}, acoustics \cite{Scienceadv2016}, cold atoms \cite{NC2019observation,aidelsburger2022cold}, etc. Therefore, conducting in-depth investigations and achieving precise control over non-Hermitian properties can provide a solid  foundation for the practical implementation of non-Hermitian physics in diverse systems.

\begin{figure}[htbp]
	\centering
	\includegraphics[width=8cm]{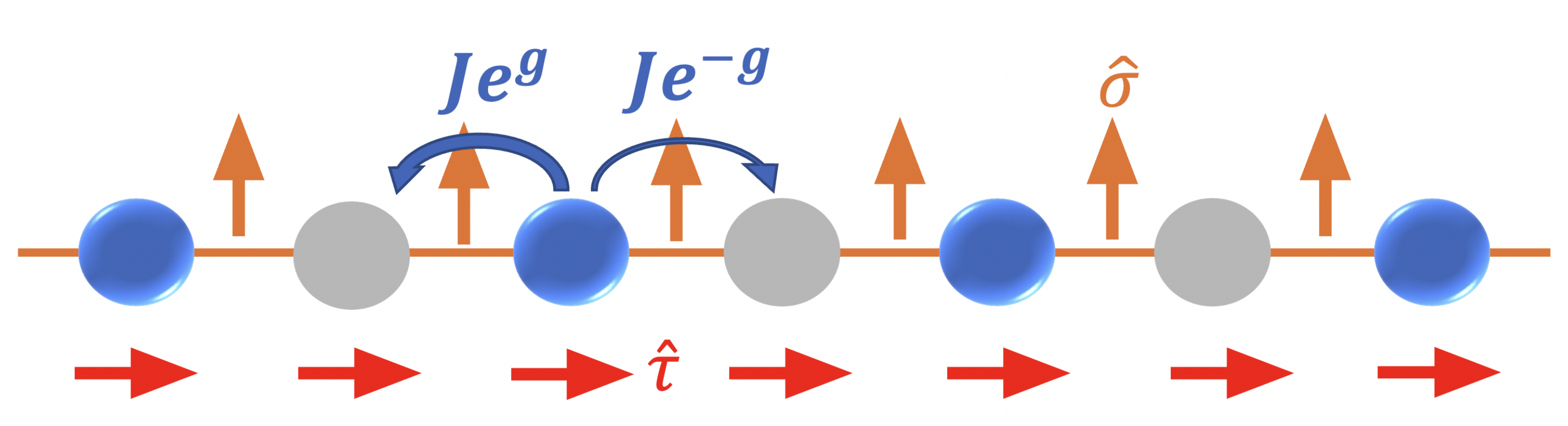}
	\caption{\label{figure1}\textbf{Schematic representation of the non-Hermitian system.} Spinless fermions (blue circles) with  nonreciprocal hopping (left-hopping amplitude $J\mathrm{e}^g$ and right-hopping amplitude  $J\mathrm{e}^{-g}$ ) are coupled via the spins  $S=1/2$ living on the bonds (orange bonds). Dual spins $\hat{\tau}$ (red arrows) are shown in the configuration corresponding to the $\hat{\sigma}$ spins (orange arrows).}
\end{figure}

Localization behaviors have been extensively investigated in the  non-Hermitian systems with nonreciprocal hopping \cite{PhysRevLett.77.570,PhysRevLett.80.2897,PhysRevLett.123.090603,PhysRevB.100.054301,PhysRevB.103.054203}, gain or loss \cite{PhysRevA.95.062118, PhysRevLett.123.090603} and quasiperiodic lattices \cite{PhysRevB.101.174205,PhysRevB.100.054301,PhysRevB.106.014204}. However, much less is known about the interplay of the non-Hermiticity and lattice gauge field. Motivated by  recent developments, the present study 
focus on  the non-Hermitian extension of  1D lattice model of spinless fermions  coupled through spins $1/2$, as illustrated in Fig.~\ref{figure1}. The spin subsystem in this model act as a  $\mathbb{Z}_2$ gauge field for the fermions. The nonreciprocal hopping introduces non-Hermiticity, thereby prompting us to investigate the nonequilibrium dynamics of systems subjected to the influence of a $\mathbb{Z}_2$ gauge field and non-Hermiticity.

Through duality mapping \cite{PhysRevD.17.2637,thesis,PhysRevB.96.205104}, we  investigate the quench dynamics of fermions in an effective  binary disorder potential. We show that the  disorder-free localization-delocalization transition arises from the interplay between the  non-Hermitian skin effects and the $\mathbb{Z}_2$ gauge field. This phase transition is closely associated with the  real-complex transition of eigenenergies. Moreover, in the dynamics of quantum mutual information of fermions and spins,  two distinct scaling laws are observed in both the localized  and delocalized phases, which are the fingerprints of the non-Hermitian  quantum disentangled liquids. In the localized phase, the $\mathbb{Z}_2$ gauge field dominates the area law scaling, while
the non-Hermitian skin effect results in the area law scaling in the delocalized phase. Furthermore, we demonstrate that the non-Hermiticity  induces the flow of quantum information  from the fermions to the spins.
 Our result not only opens a way for  preparing and controlling the quantum correlated states in the nonequilibrium systems, but also provides insights into the quantum simulation of lattice gauge theory.

\section{RESULTS}

\subsection{Model}

\begin{figure}[htbp]
	\centering
	\includegraphics[width=8.5cm]{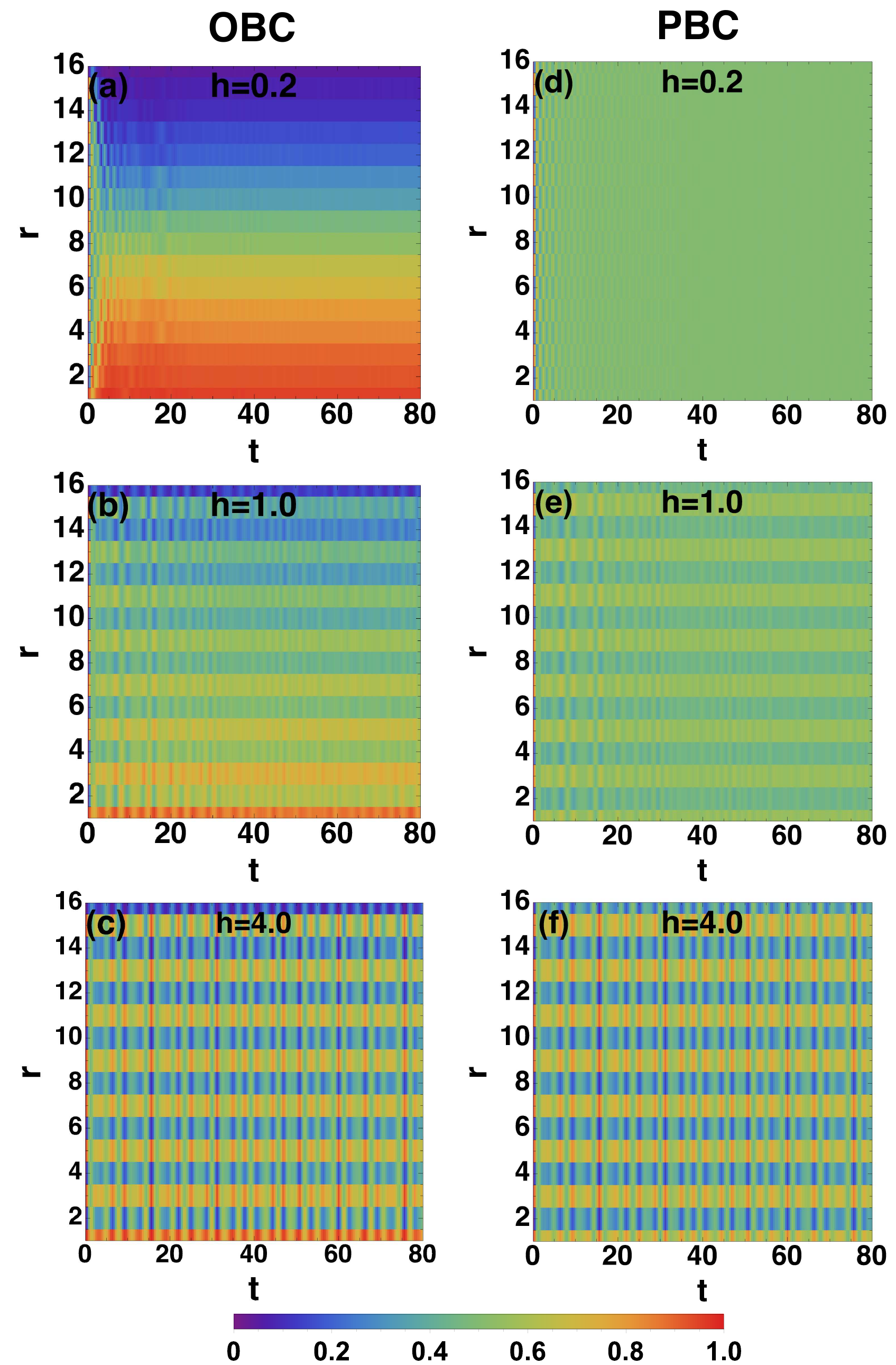}
	\caption{\label{OPBC-FD} \textbf{Dynamics of fermion density.} Dynamics of fermion density  $\left\langle n_r (t) \right\rangle$ for the system with length $L=16$ under (a)(b)(c) open boundary condition (OBC) and (d)(e)(f) periodic boundary condition (PBC) at different effective disorder strength $h$. The initial state of fermionic subsystem is charge density wave (CDW) state and the non-Hermiticity is chosen as $g=0.2$. The color bar indicates the values of fermion density.}
\end{figure}

\begin{figure*}[htbp]
	\centering
	\includegraphics[width=18cm]{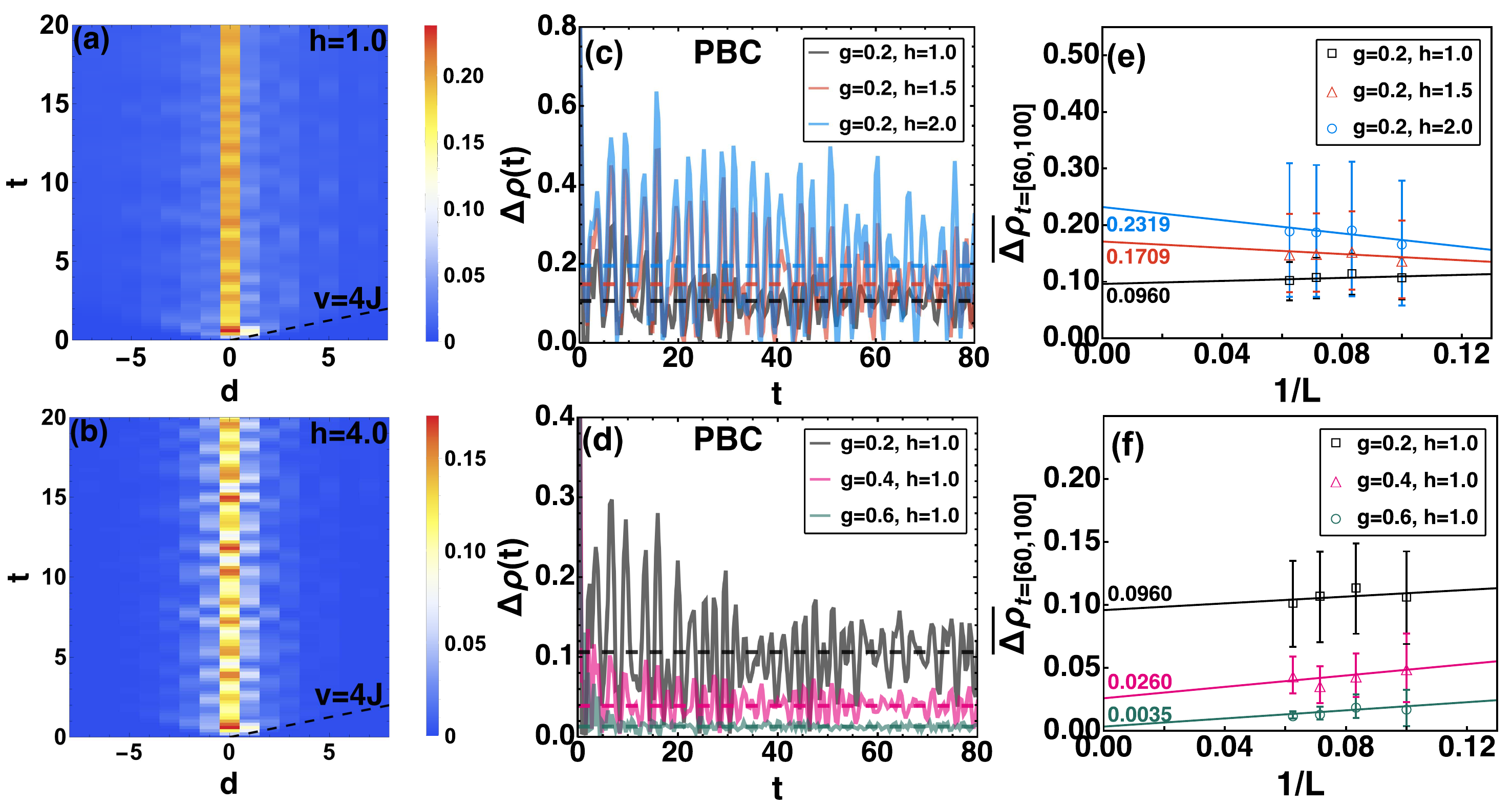}
	\caption{\label{Correlator-rhot} \textbf{Dynamics of  connected density correlator and density imbalance}. (a)(b) Absolute value  of connected density correlator $\left\langle\psi(t)\left|\hat{n}_j \hat{n}_{j+d}\right| \psi(t)\right\rangle_c $ of  fermions as a function of the separation $d$ and time $t$ for $g=0.2$.  The dashed line indicates the Lieb-Robinson velocity $v_{LR}=4J$ describing the spreading of correlations. The color bars quantify the absolute values of $\left\langle\psi(t)\left|\hat{n}_j \hat{n}_{j+d}\right| \psi(t)\right\rangle_c $. 
	Time evolution of density imbalance $\Delta \rho(t)$  after a quench for (c) a range of values $h$ at $g=0.2 $ and
		(d) a range of values $g$ at $h=1.0$.  (e)(f) Final time values of $ \overline{\Delta \rho}(t)$ averaged from $t=60$ to $t=100$ for different systems sizes up to $L=16$. Thermodynamic values for (e) different $h$ at fixed $g$ and  (f) different $g$ at fixed $h$   are labeled near the y-axis. The bars denote the standard error of the mean. All dynamics are measured   after a global quench from a charge density wave (CDW) initial state under periodic boundary condition (PBC).}
\end{figure*}
 
We consider the 1D lattice model of spinless
fermions  with the nonreciprocal hopping, where the fermions  are coupled through the bond spins characterized by $S=1/2$, see Fig.~\ref{figure1}. The Hamiltonian governing this system is given by 
\begin{eqnarray}
	\label{Hamiltonian}
	\hat{H}=&& -J \sum_{j} \hat{\sigma}_{j, j+1}^{z}\left( \mathrm{e}^g \hat{f}_{j}^{\dagger} \hat{f}_{j+1}+ \mathrm{e}^{-g}\text { H.c. }\right) \nonumber\\
	&& -h \sum_{j} \hat{\sigma}_{j-1, j}^{x} \hat{\sigma}_{j, j+1}^{x}
\end{eqnarray}
where   $J$ and $h$ are the fermion tunnelling strength and Ising coupling, respectively. $J\mathrm{e}^g$ ($J\mathrm{e}^{-g}$) is the left (right)-hopping amplitude introducing   the  non-Hermiticity which could be realized in synthetic matter using, for example, photonic systems \cite{S2020topological,S2021generating}, topolectrical circuits \cite{NP2020generalized}, or  ultracold atom systems \cite{PhysRevX.8.031079, NC2019observation}. In the
following, we will set $J = 1$ as the units of energy and choose a fixed fermion number $N_\mathrm{f}=L/2$ (half-filling), where $L$ is the number of sites.
This system is invariant under the ${\mathbb{Z}}_{2}$  gauge transformation
\begin{eqnarray}
	\hat{f}_i \rightarrow \gamma_i  \hat{f}_i, \quad \hat{f}^\dagger_i \rightarrow \gamma_i  \hat{f}^\dagger_i, \quad \hat{\sigma}_{j, j+1}^{z} \rightarrow \gamma_i \hat{\sigma}_{j, j+1}^{z} \gamma_{i+1},
\end{eqnarray}
where $\gamma_i = \pm 1$ is a local gauge function. The generator of the  ${\mathbb{Z}}_{2}$  gauge transformation is given by the unitary operator
\begin{eqnarray}
	G=\prod_i G_i^{(1-\gamma_i)/2}, 
\end{eqnarray}
where $G_i =(-1)^{\hat{n}_i} \hat{\sigma}_{i-1, i}^{z}  \hat{\sigma}_{i, i+1}^{z}$, and $\hat{n}_i=\hat{f}^\dagger_i \hat{f}_i $ is the local number operator of fermions. It is note that $G_i^2 =1$ and $\left[G_i, G_j\right]=0$ for all $i,j$. Therefore, the Hamiltonian is gauge invariant, i.e., $\left[G,H\right]=0$ for all choices of gauge function $\gamma_i$.

In order to  explore the dynamics of fermions, an extensive number of conserved charges can be identified by a duality mapping (see Methods~\ref{methods}). Then an effective Hamiltonian with fermion operator  $\hat{C}_j =\hat{\tau}_j \hat{f}_j$ is written as
\begin{eqnarray}
	\label{Hamiltonian2}
	\hat{H}_{\left\lbrace q_i\right\rbrace}=&&-J\sum_i (\mathrm{e}^g \hat{C}_i^{\dagger} \hat{C}_{i+1} +\mathrm{e}^{-g} \hat{C}_{i+1}^{\dagger}\hat{C}_{i}) \nonumber\\
	&& +2h\sum_i \hat{q}_i (\hat{C}_i^{\dagger}\hat{C}_{i}-1/2),
\end{eqnarray}
which describes a nonreciprocal hopping tight-binding model with the binary disorder potential dominated by the charge sectors $\left\lbrace q_j \right\rbrace $.

We choose the tensor products of fermion and spin states $|\Psi_0\rangle =  \left|S \right>_{\mathrm{\sigma}} \otimes \left|\psi \right\rangle_\mathrm{f} $ as the initial state that  can be transformed to the one in the Hilbert space of the  $\hat{C}$ degrees of freedom,
\begin{eqnarray}
\label{initial state}
|\Psi_0\rangle &&=|\uparrow \uparrow \uparrow \cdots\rangle_{\sigma} \otimes \left|\psi\right\rangle_\mathrm{C} \nonumber \\
 &&= \frac{1}{\sqrt{2^{N-1}}}
\sum_{\lbrace q_i \rbrace=\pm 1} \left|q_1, q_2, \cdots, q_N \right\rangle \otimes \left|\psi\right\rangle_\mathrm{C},
\end{eqnarray}
where the spin states are restricted to the z-polarized state  $|S\rangle_{\sigma}=|\uparrow \uparrow \uparrow \cdots\rangle$, and the sum is over all possible charge configurations  $\left\{ q_i\right\}=\pm 1$.
The quench dynamics of whole system is governed by
\begin{equation}
|\Psi(t)\rangle =\frac{\mathrm{e}^{-\mathrm{i} \hat{H}t} |\Psi_0\rangle}{\Vert \mathrm{e}^{-\mathrm{i} \hat{H}t} |\Psi_0\rangle \Vert},
\end{equation}
 where the normalization takes into account the fact that  the norm is not conserved due to the lack of unitary for the non-Hermitian systems. Here, the non-Hermitian dynamics  is characterized by the individual quantum trajectories  without quantum jumps, which describes physics differently from the Lindblad master equation approach where the quantum jumps occur and the outcomes are averaged out \cite{PhysRevLett.123.090603}.


\subsection{Non-Hermitian skin effect}

\begin{figure*}[htbp]
	\centering
	\includegraphics[width=18cm]{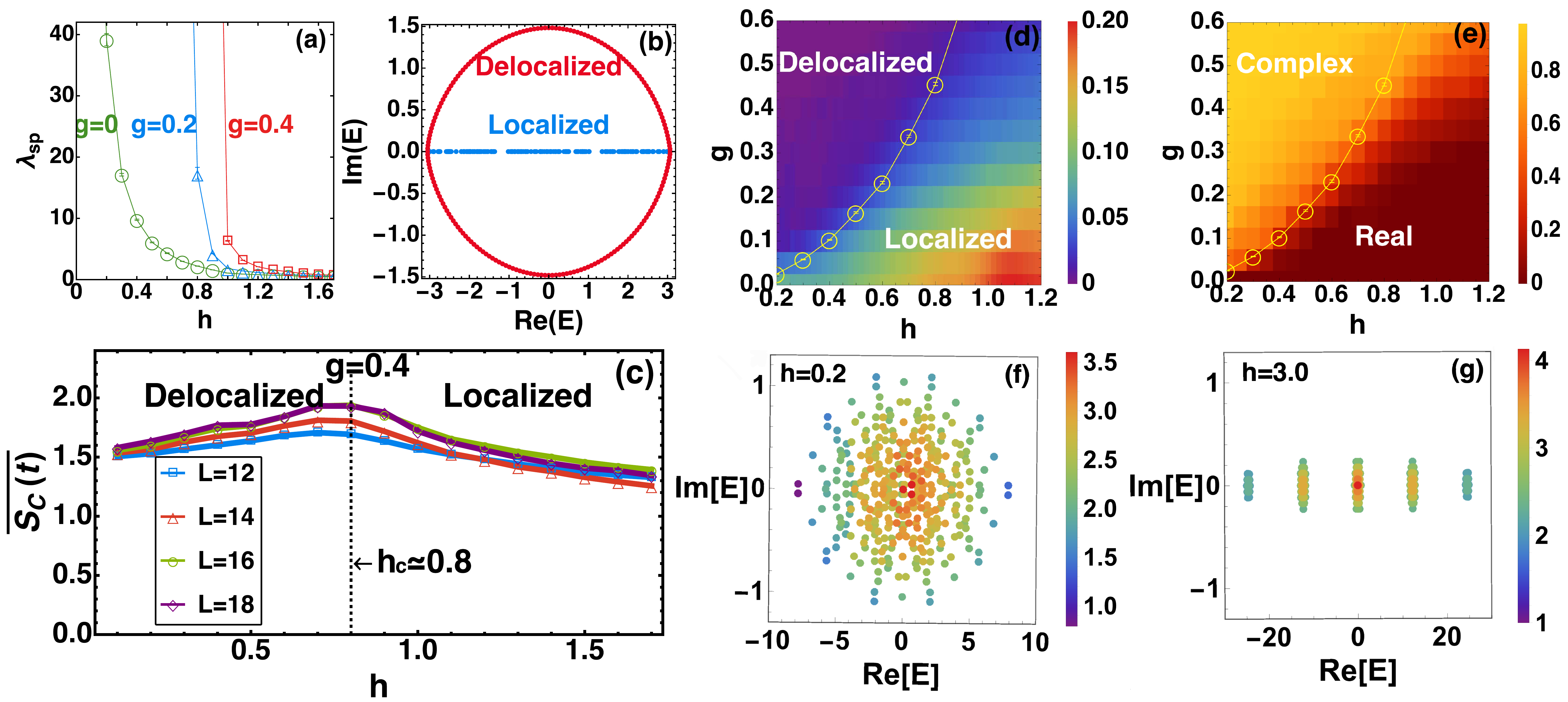}
	\caption{\label{transition}  \textbf{Phase transitions}. (a) The single particle localization lengths are obtained from the transfer matrix method as a function of $h$ for the system with $L=200$. 
	The bars indicate the  statistical errors on  the mean of random selection of $2000$ charge sectors. (b) The eigenenergies of the single-particle model with $L=200$ for the localized phase $g=0.1, h=0.6$ and delocalized phase $g=0.8, h=0.6$. (c) Final time value of entanglement entropy of fermions is averaged from $t = 500 $ to $t = 1000$ for different $L$, and the entanglement entropy is obtained by averaging over $1000$ charge sectors. (d) Phase diagram of localization-delocalization transition obtained by mapping the averaged  values of $\Delta \rho(t)$ (from $t=120$ to $t=160$) to the  parameter space $(g,h)$ for  $L=12$ with all charge sectors. The delocalization transition lines are displayed from the relation $g_\mathrm{c}=\lambda_\mathrm{sp}^{-1} $, where $\lambda_\mathrm{sp}$ is the single-particle localization length. The color bar quantifies the the averaged  values of $\Delta \rho(t)$. (e) Dependence of $P_{\mathrm{Im}}$ on $(h,g)$ for the single-particle model with $L=200$, the color bar quantifies the values of $P_{\mathrm{Im}}$ .  Eigenenergies of the non-Hermitian Hamiltonian Eq.~(\ref{Hamiltonian2}) with $L=12$  for a fixed sector $\left\{ 1, -1, 1, -1,\dots,1,-1 \right\}$ in the (f) delocalized phase $h=0.2$, $g=0.2$ and (g)  localized phase $h=3.0$, $g=0.2$. The color of  each circle   represent the intensity of entanglement entropy of  right eigenstate corresponding to the eigenenergy.}
\end{figure*}

We first consider the initial state of fermions to be of a charge density wave (CDW) state, and  monitor the dynamics of the fermion density $\left\langle n_r (t) \right\rangle = \left\langle \Psi(t) \right| \hat{n}_r |\Psi(t)\rangle $.  In the Hermitian case,  fermions propagate reciprocally throughout  the system after the global quench. However, when the hopping is nonreciprocal, fermions can be amplified toward one direction and attenuated in the other direction.
For the open boundary condition (OBC), see Figs.~\ref{OPBC-FD}(a)(b)(c), when $h=0.2$,
the memory of initial CDW state is lost after the global quench. The fermions propagate to one side leading to the non-Hermitian skin effect, and a approximate domain wall state is formed due to the Pauli exclusion principle. As $h$ is increased to $0.6$, a little memory of initial state is  preserved because of  the effective binary disorder potential,  and the localization phenomenon is consistence with the non-Hermitian skin effect.
Even when $h$ is as large  as $4.0$,  the trail of  non-Hermitian skin effect is still observed. Meanwhile, the large fluctuations are  also observed on account of the binary nature of the disorder. In the periodic boundary condition (PBC) case,  see Figs.~\ref{OPBC-FD}(d)(e)(f), the non-Hermitian skin effect disappears, and the localization behavior and fluctuations become more pronounced with  the  increase of the effective disorder strength $h$. We also consider the quench dynamics from the initial domain wall (DW) state, the results about non-Hermitian skin effects are shown in Supplementary Note 1. In addition, we also present the dynamics of spin subsystem arising from the CDW and DW  initial states  in Supplementary Note 2.

\subsection{Disorder-free localization-delocalization transition}
The Anderson model with the nonreciprocal hopping, which has been  investigated by Hatano and Nelson \cite{PhysRevLett.77.570}, exhibits the  localization-delocalization phase transition in one-dimensional system. We show that  the  interplay between the $\mathbb{Z}_2$ gauge field and non-Hermitian skin effect lead to the
localization-delocalization phase transition without disorder.

The light-cone behavior exhibited by the connected density correlator   is   an important signal for diagnosing the localization phenomenon. The connected density correlator for fermions is defined by
\begin{eqnarray}
&&\left\langle\Psi(t)\left|\hat{n}_j \hat{n}_{j+d}\right| \Psi(t)\right\rangle_\mathrm{c} =\left\langle\Psi(t)\left|\hat{n}_j \hat{n}_{j+d}\right| \Psi(t)\right\rangle - \nonumber \\
&&\left\langle\Psi(t)\left|\hat{n}_j\right| \Psi(t)\right\rangle\left\langle\Psi(t)\left|\hat{n}_{j+d}\right| \Psi(t)\right\rangle,
\end{eqnarray}
where $d$ is the separation between two sites.
In Figs.~\ref{Correlator-rhot}(a)(b), we  observe
the asymmetric  light-cone behavior in the evolution but  for the strong effective disorder the spreading eventually halts. Particularly,
the spreading of correlations is well described by the  Lieb-Robinson velocity $v=4J$, indicating that the quantum information propagates with a finite velocity \citep{PhysRevLett.97.050401,Essler_2016}.
Next, we probe the localization behavior by measuring the average of nearest-neighbour density imbalance,
\begin{equation}
	\Delta \rho(t)=\frac{1}{L} \sum_{j}\left|\left\langle \Psi(t) \left|\hat{n}_{j}-\hat{n}_{j+1}\right| \Psi(t)\right\rangle\right|,
\end{equation}
which quantifies the average difference in fermion density between nearest-neighbour sites. For a translationally invariant state, this quantity is zero, while in the localized phase, it takes a non-zero value.  From  Figs.~\ref{Correlator-rhot}(c)(d), we can  see that the  $\Delta \rho(t)$ decays  to a finite asymptotic value with some fluctuations. The asymptotic value of $\Delta \rho(t)$ and oscillation amplitude both grow   with  the increase of $h$ and decreases as $g$ increases. In the finite-size scaling displayed in Figs.~\ref{Correlator-rhot}(e)(f),  the asymptotic value of $\Delta \rho(t)$ approaches the   thermodynamic value  as $L$ becomes larger. The strong disorder strength prevents the delocalization, while the strong non-Hermiticity results in the delocalization, indicating the weak memory effects.

Von Neumann entanglement entropy  has been widely used to study the proprieties of many-body systems. Its definition is given by
 \begin{eqnarray}
 	S=-\mathrm{Tr}\left[\varrho^\mathrm{A} \ln \varrho^\mathrm{A}\right],
 \end{eqnarray}
where the reduced density matrix $\varrho^\mathrm{A}$ is the partial trace of the density matrix $\varrho$ of whole system  $\varrho^\mathrm{A}=Tr^\mathrm{B}\left[\varrho\right]$. If $\mathrm{A}$ and $\mathrm{B}$ are entangled, then the reduced density matrix must be a mixed state and the von Neumann entanglement entropy measures this mixing. In Fig.~\ref{transition}(c), the entanglement entropy of nonequilibrium steady state reaches the maximum at the critical point, and  as the length increases the critical point $h_\mathrm{c} \simeq 0.8 $  is approached, separating the delocalized and localized phases. More results about the dynamics of entanglement entropy of fermion can be found in Supplementary Note 3.

The aforementioned results provide insight into the impact of the interplay between non-Hermiticity and ${\mathbb{Z}}_{2}$ gauge field on the phenomenon of localization-delocalization transition.
The dynamical phase paragram is shown in  Fig.~\ref{transition}(d) where the finite final time values of $\Delta \rho(t)$ corresponds to the localized phase. The delocalization line is determined by the  localization length that characterizes the localization features. Since the effective Hamiltonian Eq.(\ref{Hamiltonian2}) describes the noninteracting fermions with a binary disorder potential, the localization length of single-particle model also reveals the localization feature.
  In the non-Hermitian case, $g$ plays the role of an imaginary vector potential that appears in the wave functions of a localized state such that \cite{PhysRevB.105.024303},
  \begin{eqnarray}
  \psi^\mathrm{L, R}(r) \sim \exp \left(-\frac{\left|r-r_\mathrm{c}\right|}{\lambda} \mp g\left(r-r_\mathrm{c}\right)\right),
  \end{eqnarray}
  where $\psi^\mathrm{L, R}(r)$ are left and right eigenvectors, respectively. $r_\mathrm{c}$ is its localization center and $\lambda$ represents the corresponding localization length. If $g > \lambda^{-1}$, either $\psi^{\mathrm{L}}(r)$ or $\psi^{\mathrm{R}}(r)$ diverges, the wave functions $\psi^{\mathrm{L, R}}(r)$ no longer represent an exponentially localized state. Therefore, the delocalized transition point is determined by the condition  $g=\lambda^{-1}$ \cite{PhysRevB.56.R4333,PhysRevB.56.8651,PhysRevB.103.064201,PhysRevB.105.024303}. For $g<\lambda^{-1}$, the energy spectrum is real and all wave functions are exponentially localized with the same localization length, so that the spectral localization also implies dynamical localization \cite{PhysRevLett.75.117,PhysRevB.103.054203}. In Fig.~\ref{transition}(a), the single particle  localization length, obtained by the  transfer matrix method (see Supplementary Note 4) \cite{BKramer_1993,PhysRevLett.126.166801,PhysRevB.104.104203}, decreases exponentially with the increase of $h$. The localization length  diverges at a critical point $h_\mathrm{c}$  and the larger $g$ induces the larger $h_\mathrm{c}$. As shown in Fig.~\ref{transition}(d), the transition line   $g_\mathrm{c}=\lambda_{\mathrm{sp}}^{-1}$  clearly  distinguishes the  localized  and delocalized phases.

  \begin{figure*}[htbp]
	\centering
	\includegraphics[width=18cm]{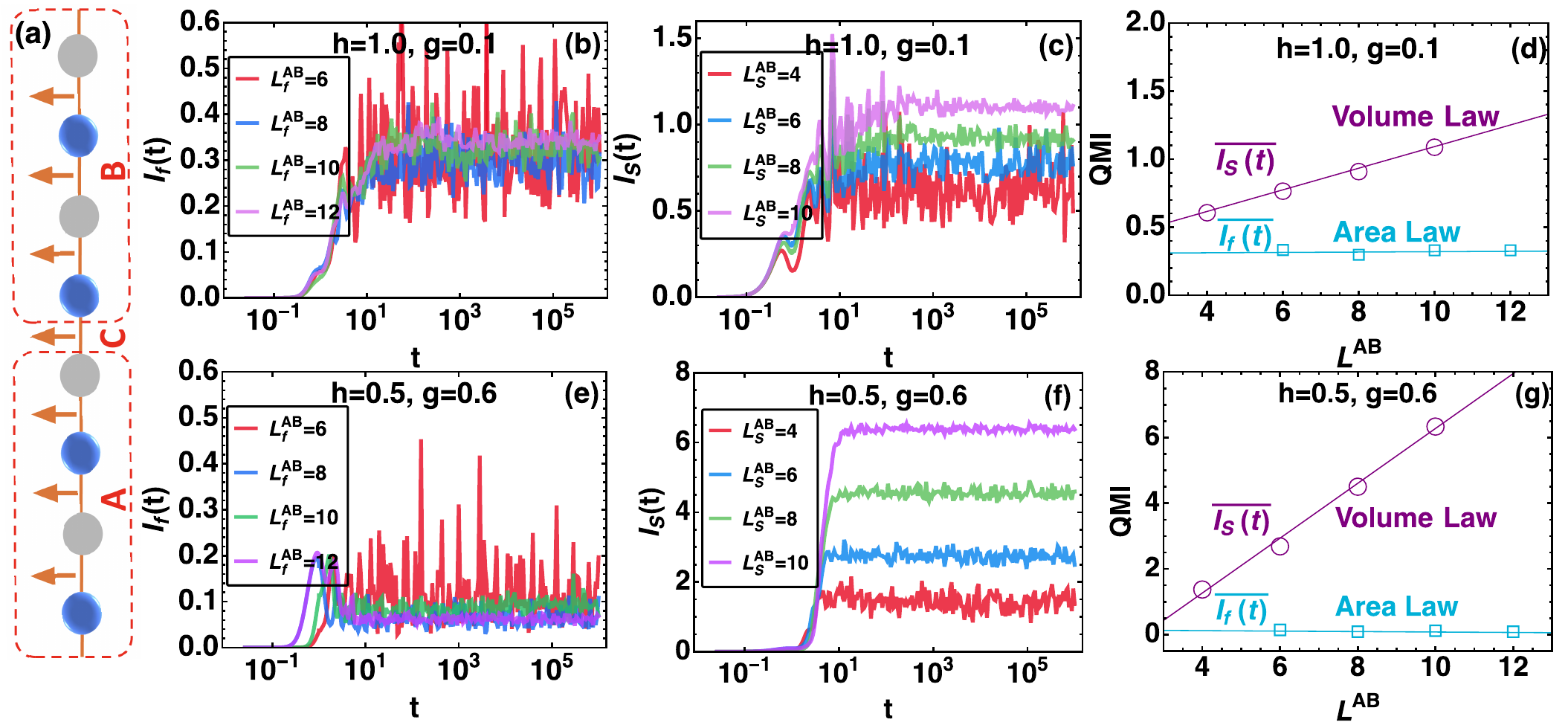}
	\caption{\label{mutual-scaling} \textbf{Features of non-Hermitian quantum disentangled liquid.} (a) The partitioning methods of the whole system.  Dynamics of quantum mutual information for the fermion and spin subsystems with different sizes at the localized phase (b)(c)(d) $h=1.0, g=0.1$ and delocalized phase (e)(f)(g) $h=0.5, g=0.6$. The initial state  is the charge density wave  state for fermion and  the $z$-polarized  state for spins. Saturation values of quantum mutual information of fermion $ \overline{I_\mathrm{f} (t)}$ and spin $ \overline{I_\mathrm{S} (t)}$ are averaged from $t=10^3$ to $t=10^6$ for different systems sizes. Here the open boundary condition  is considered. }
\end{figure*}

\subsection{Real-complex transition}
It has been discovered that the localization-delocalization transition is accompanied by a real-complex transition of the eigenenergies in the non-Hermitian system \cite{PhysRevLett.123.090603}. In order to verify this phenomenon in a disorder-free system,  we will examine it utilizing the ratio
\begin{eqnarray}
   P_{\mathrm{Im}}=\overline{D_{\mathrm{Im}}/D},
 \end{eqnarray}
 where $D_{\mathrm{Im}}$ is the number of eigenvalues  with nonzero imaginary parts and $D$ is the total number of eigenvalues. The overline notation indicates the disorder average. We have employed a cutoff of $C=10^{-13}$ in the calculation.  In Fig.~\ref{transition}(b), the spectrum is symmetric around the real axis due to the time-reversal symmetry. In the delocalized phase,  the imaginary parts of eigenvalues  are nonzero, whereas in the localized phase, they are zero. Fig.~\ref{transition}(e) depicts the phase diagram of real-complex transition  which is approximately coincide with the localization-delocalization transition. In the localized phase, almost all the eigenenergies are real due to the strong disorder, while in the delocalized phase they are complex  due to the nonreciprocal hopping. The non-Hermiticity  leads to the dynamical instability in the delocalized phase. In contrast, the system is stable in the localized phase where the energy is conserved.  Consequently, the dynamics are distinct in the two phases,  the dynamical localization-delocalization transition closely corresponding to the real-complex transition of eigenenergies.

	\begin{figure*}[ht]
		\centering
		\includegraphics[width=18cm]{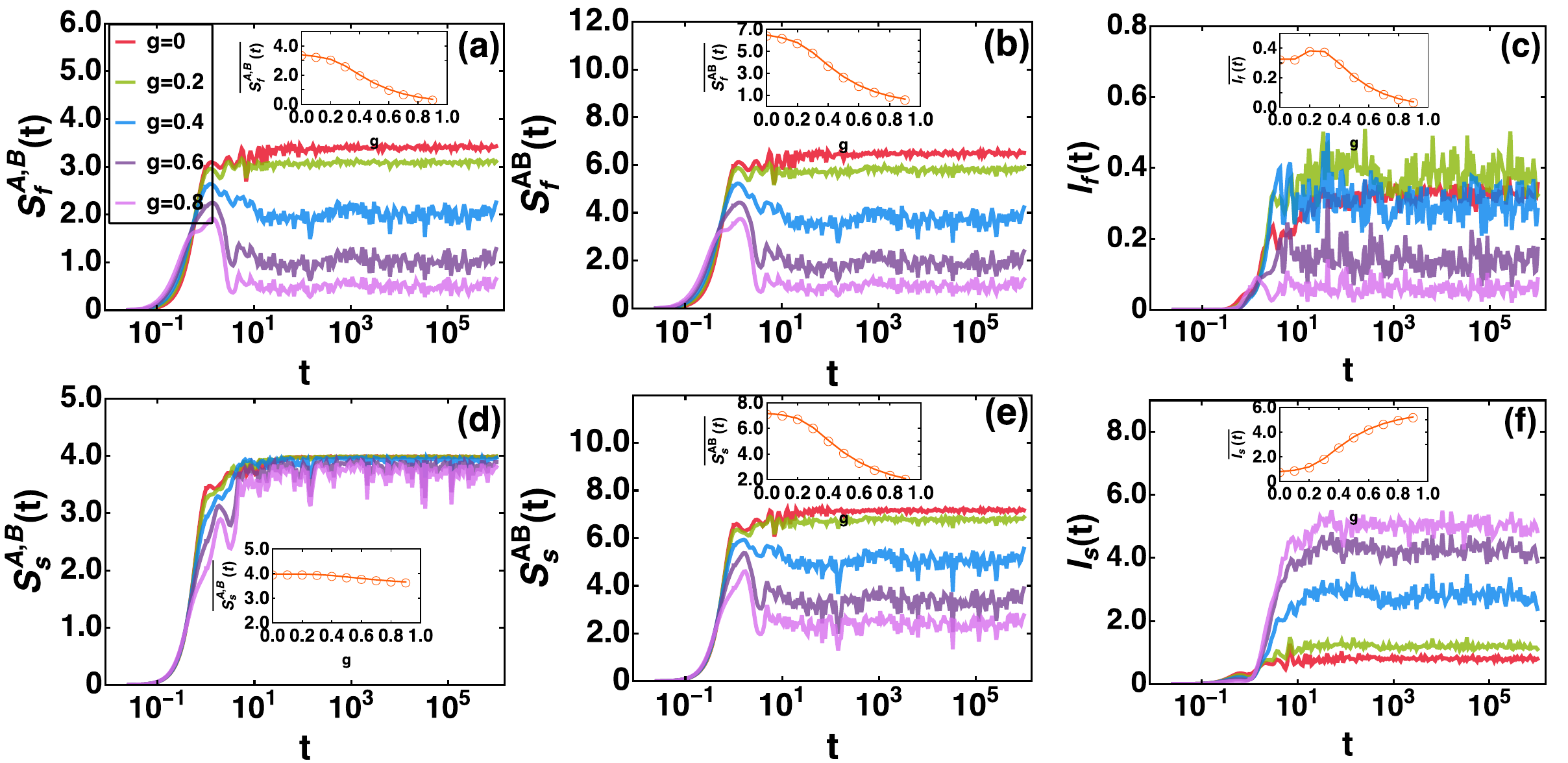}
	\caption{ \label{mutual-h1} \textbf{Dynamics of entanglement entropy and quantum mutual information.} Dynamics of entanglement entropy (a) $S^\mathrm{A}_\mathrm{f} (t)$,   $S^\mathrm{B}_\mathrm{f} (t)$, (b) $S^\mathrm{AB}_\mathrm{f} (t)$ and (c) quantum mutual information $I_\mathrm{f} (t)$ of  fermion subsystems  after a quench from a charge density wave initial state under open boundary condition. Dynamics of entanglement entropy (d) $S^\mathrm{A}_\mathrm{S} (t)$, $S^\mathrm{B}_\mathrm{S} (t)$, (e) $S^\mathrm{AB}_\mathrm{S} (t)$ and (f)  quantum mutual information of spin subsystems  after a quench from a charge density wave   initial state under open boundary condition. Insets are the variation of mean values from $t=10^3 $ to $t= 10^6$ for the nonreciprocal hopping strength $g$.   All results are obtained under the same conditions $L=10$ and $h=1$. }
	\end{figure*}

In Figs.~\ref{transition}(f)(g), it can be observed that the strong disorder not only suppresses the generation of   imaginary parts in almost all eigenenergies, but also enlarges the absolute value of real parts of eigenenergies and leads to the  formation of eigenenergy clusters, attributable to the binary nature of the disorder.
More intriguingly, the right eigenstates corresponding to the  small absolute real or imaginary parts of eigenenergies tend to have larger entanglement entropy regardless of whether they belong to the delocalized or localized phases.

\subsection{Non-Hermitian quantum disentangled liquids}
 The interacting mixture of heavy and light particles has been recognized as a factor that can  induce a dynamical form of localization, where the heavy particles act as a source of disorder for the light ones. From the perspective of  entanglement, the multiple subsystems consisting of the heavy and light particles may behave diverse dynamics, with some undergoing thermalization, while others do not, which is described by  a phase of matter known as  quantum disentangled liquid \citep{Grover_2014,PhysRevLett.119.176601,thesis}. It is natural to inquire whether the quantum disentangled liquid can persist in the non-Hermitian system. In order to explore this problem further, we will  reveal the long-time quench dynamics of quantum mutual information of  the original degrees of freedom, namely the $f$-fermions and $S$-spins.  For the OBC case,  we divide the system into three partitions, as depicted in Fig.~\ref{mutual-scaling}(a), to ensure equal  sizes for the two subsystems $\mathrm{A}$ and $\mathrm{B}$. The quantum mutual information between subsystems $\mathrm{A}$ and $\mathrm{B}$ is  defined as
\begin{eqnarray}
	I(\mathrm{A}:\mathrm{B})=S^\mathrm{A} +S^\mathrm{B} -S^\mathrm{AB}
\end{eqnarray}
where $S^\mathrm{A}$ and $S^\mathrm{B}$ are the von Neumann entanglement entropy for $\mathrm{A}$ and $\mathrm{B}$, respectively. The $S^\mathrm{AB}$ denotes the  von Neumann entanglement entropy of composite system $\mathrm{A}\bigcup \mathrm{B}$. The presence of  non-zero quantum mutual information indicates a correlation between the subsystems $\mathrm{A}$ and $\mathrm{B}$.

In the localized phase (see Figs.~\ref{mutual-scaling}(b)(c)(d)), it is observed that  the quantum mutual information of nonequilibrium steady states of fermions remains independent of the system size, reaching a saturation point governed by the area law, $I_\mathrm{f} (t\rightarrow\infty) \sim L_\mathrm{AB}^{\mathrm{dim}-1}$.  This behavior can be attributed to the strong  ${\mathbb{Z}}_{2}$  gauge field, which causes the fermion wave function to resemble a product state with low entanglement, indicating  that the fermions are disentangled. In contrast, the long-time dynamics of quantum mutual information of spins  depends on the system size and follows a volume law scaling,  $I_\mathrm{S} (t\rightarrow\infty) \sim L_\mathrm{AB}^{\mathrm{dim}}$, as shown in Fig.~\ref{mutual-scaling}(c). This distinct behavior, where the light particles (fermions) fail to thermalize and reside in a disentangled area law phase due to the presence of the heavy particles (spins),  serves as a prominent characteristic of quantum disentangled liquid.

In the delocalized phase,  as illustrated in Figs.~\ref{mutual-scaling}(e)(f)(g),  the  fermions and spins respectively exhibit the area law scaling and volume law scaling,   which implies that the quantum disentangled liquid  retains despite the strong non-Hermiticity.
This result challenges the conventional understanding that the delocalized phase is characterized by the volume law scaling of quantum correlation.   In the closed quantum systems, the particles diffuse throughout the system during evolution, resulting in the thermalization and extensive quantum correlation. However, the non-Hermitian skin effect drives the fermions to propagate to one side and preventing the diffusion of quantum correlation  throughout the system. As a result, the fermions are delocalized but not thermalized, and cannot generate the global quantum correlation \cite{PhysRevX.13.021007}. Therefore, the quantum mutual information of fermions  obeys the area law rather than the volume law, indicating the presence of  quantum disentangled liquid in this delocalized phase.

Although the quantum disentangled liquids  exist in both the localized and delocalized phases, their physical essences are completely different. The area law scaling for fermions in localized phase originates from the  $\mathbb{Z}_2$ gauge field, while in the delocalized phase, it is the consequence of non-Hermitian skin effect. In general, the quantum disentangled liquid shows its robustness to the non-Hermiticity. It is worth emphasizing that  this phenomenon is independent of the segmentation method used for the whole system. Even if the sizes of segments $\mathrm{A}$ and $\mathrm{B}$ are not equal, it is still observable (see Supplementary Note 5). Especially, the area law of quantum mutual information has recently been experimentally verified  in an ultracold atom simulator  \cite{NP2023verification}. Therefore,  the non-Hermitian quantum disentangled liquid may be  observed in the ultracold atom simulator through the diverse scaling behaviors of quantum mutual information.

\subsection{Flow of quantum information}

 We further study the dynamics of entanglement entropy of fermion and spin  subsystems to explore the effect of non-Hermiticity on the quantum mutual information. In the presence of non-Hermiticity, as shown in Figs.~\ref{mutual-h1}(a)(b), the growth of the entanglement entropy $S^\mathrm{A}_\mathrm{f} (t)$,   $S^\mathrm{B}_\mathrm{f} (t)$ and $S^\mathrm{AB}_\mathrm{f} (t)$ of fermion is greatly suppressed. The entanglement entropy for the nonequilibrium steady state is  smaller than that for the Hermitian case and monotonically decreases with respect to  $g$.   Consequently, the quantum mutual information of fermion (see Fig.~\ref{mutual-h1}(c)) initially increases and then decreases as a function of $g$. Fig.~\ref{mutual-h1}(d) shows that the non-Hermiticity has minimal  effects on the dynamics of entanglement entropy of spin subsystems $S^\mathrm{A}_\mathrm{S} (t)$ and $S^\mathrm{B}_\mathrm{S} (t)$, however, the presence of non-Hermiticity significantly diminishes the entanglement entropy of nonequilibrium steady state (see Fig.~\ref{mutual-h1}(e)),
 which are the mainly reason for the growth of quantum mutual information with the non-Hermiticity, see Fig.~\ref{mutual-h1}(f). The non-Hermiticity induces the nonreciprocal dissipation of fermion system, causing the quantum information to flow from fermions  to spins. Consequently,  the quantum mutual information   of spin subsystem  increases obviously. This result paves the way for the  preparation and control of the quantum correlated states in the non-Hermitian systems operating far from thermal equilibrium.


\section{Discussion}

In summary, we have investigated the nonequilibrium dynamics in the non-Hermitian lattice gauge model  where the spinless fermion coupled to a $\mathbb{Z}_2$ gauge field formed by the spins on the bonds. We show that the interplay between the non-Hermitian skin effect and  the $\mathbb{Z}_2$ gauge field leads to the disorder-free localization-delocalization transition. 
We have also explored the  real-complex transition of eigenenergies  which closely corresponds to the  localization-delocalization transition. Additionally,  by identifying the diverse scaling behaviors of quantum mutual information where the fermions obey the area law and the spins follow the volume law, we have proposed a  phase of matter known as the non-Hermitian quantum disentangled liquids, and demonstrated the existence of these disentangled liquids in both   the localized and delocalized phases.
The area law scaling of fermions in the localized phase is due to the $\mathbb{Z}_2$ gauge field, whereas  in delocalized phases it originates from the non-Hermitian skin effect. 
Moreover, the  nonreciprocal dissipation of fermions results in  the flow of the  quantum information  from the fermions to the spins, and induces the growth of quantum mutual information of spin subsystem. 

It is worth noting that previous discussions have focused on the localization behavior and quantum disentangled liquid in the Hermitian case ($g=0$) \cite{PhysRevLett.118.266601,PhysRevLett.119.176601,thesis}. However, we want to emphasize  that the introduction of non-Hermiticity leads to  several intriguing phenomena  that are absent in the Hermitian case, such as  the non-Hermitian skin effect,  the disorder-free localization-delocalization transition and  the real-complex transition of eigenenergies. In particular, the non-Hermitian skin effect gives rise to the  non-Hermitian quantum disentangled liquid in the delocalized phase, which is completely different from the Hermitian case where the $\mathbb{Z}_2$ gauge field dominates the generation of quantum disentangled liquid. Furthermore, the flow of quantum information from the fermions to the spins, induced by non-Hermiticity, has not been previously discussed.

 Experimentally, the implementation of ${\mathbb{Z}}_{2}$ lattice gauge fields  has been achieved using  current technological capabilities \cite{martinez2016real,schweizer2019,NP2019realization,scienceadvance2019,PhysRevResearch.4.L022060,PhysRevX.10.021041}, including the coupling of bosons and fermions to the gauge field.   These experiments offer the high controllability and
precise readout,  providing a toolbox to realize lots of quantum simulation experiments. In our study, the  non-Hermitian system  may be experimentally realized  using  the ultracold atoms in optical lattices, where the coupling of fermions to a ${\mathbb{Z}}_{2}$ lattice gauge field is to use the Floquet scheme \cite{NP2019realization,schweizer2019}, and the non-Hermiticity is introduced by the cold atoms with engineered dissipation \cite{PhysRevX.8.031079,NC2019observation}.
 Recently, the superconducting circuit has been applied to investigate the  emergent ${\mathbb{Z}}_{2}$  gauge invariance  \cite{PhysRevResearch.4.L022060} and the quantum evolution under the non-Hermitian Hamiltonians \cite{dogra2021quantum}, which may provide another platform to simulate the non-Hermitian lattice gauge theory. In addition, 
 it is also noteworthy that the  electric circuits possess the capability to simulate the $U(1)$ lattice gauge theory \cite{PhysRevB.105.205141} and nonreciprocal non-Hermitian system \cite{ezawa2019non,Research2021}, thereby offering a  potential  alternative for realizing the simulation of fermionic matter coupled to the ${\mathbb{Z}}_{2}$ lattice gauge field within a non-Hermitian system.  Collectively, these experimental platforms such as cold atoms, superconducting circuits, and electric circuits separately  offer distinctive avenues for exploring the lattice gauge fields and non-Hermitian systems.
  However, a crucial challenge remains in fully  implementing  the  non-Hermitian lattice gauge theories on these experimental platforms. 
  The main obstacle lies in the fact that  the lattice gauge theories, which describes the coupling of matter fields to gauge fields, primarily arise in interacting systems, while the non-Hermitian Hamiltonians have thus far  been implemented in the context of few-body  or noninteracting systems \cite{PhysRevX.8.031079,NC2019observation,dogra2021quantum}.
  Consequently,  the experimental introduction of non-Hermiticity into the lattice gauge theories is still an unresolved challenge in terms of the current experimental techniques. Nevertheless, given the rapid advancement in quantum simulation technologies and the increasing adoption of diverse platforms, it might happen that the non-Hermitian lattice gauge theory can be effectively realized in forthcoming experiments.
Recently, the many-body non-Hermitian skin effect has been theoretically investigated in an atom-cavity hybrid system where fermionic atoms in the one-dimensional lattice are 
subject to a cavity-induced dynamic gauge potential \cite{li2023manybody}, which provides  another  possible  scheme to test and verify our results in the interacting system.
Furthermore, the advancements in measurement technologies make it possible to experientially  access the quantum  entanglement and quantum mutual information \cite{N2015measuring,Lukin256,NP2023verification}.
Once experimental systems closely matching our model  can be synthesized, our results will provide valuable insights for interpreting  the quantum simulation experiments that probe the disorder-free localization properties and non-Hermitian quantum disentangled liquid.

\section{Methods\label{methods}}

In this section, we outline the procedure of duality mapping. Through duality mapping from bond spin $\hat{\sigma}$ to site spins $\hat{\tau}$,

\begin{equation}
	\hat{\tau}_{j}^{z}=\hat{\sigma}_{j-1, j}^{x} \hat{\sigma}_{j, j+1}^{x}, \quad \hat{\sigma}_{j, j+1}^{z}=\hat{\tau}_{j}^{x} \hat{\tau}_{j+1}^{x}
\end{equation}
 the Hamiltonian Eq.~(\ref{Hamiltonian})  becomes
 \begin{eqnarray}
 	\label{Hamiltonian1}
 	\hat{H}^{\prime}=&&-J \sum_{i} \left( \mathrm{e}^g \hat{\tau}_{i}^{x} \hat{\tau}_{i+1}^{x} \hat{f}_{i}^{\dagger} \hat{f}_{i+1} + \mathrm{e}^{-g} \hat{\tau}_{i}^{x} \hat{\tau}_{i+1}^{x} \hat{f}_{i+1}^{\dagger} \hat{f}_{i} \right) \nonumber\\
 	&&-h \sum_{i} \hat{\tau}_{i}^{z},
 \end{eqnarray}
which  is equivalent to the original one only on a restricted Hilbert space. It can be identified  that the charges $\hat{q}_{j}=\hat{\tau}_{j}^{z}(-1)^{\hat{n}_{j}}$ with $\hat{n}_j=\hat{f}^\dagger_j \hat{f}_j$ are local conserved and  generate the ${\mathbb{Z}}_{2}$ gauge symmetry.
 By introducing a  spinless fermion operators $\hat{C}_j =\hat{\tau}_j \hat{f}_j$, the effective Hamiltonian Eq.~(\ref{Hamiltonian2}) can be obtained.

We choose the tensor products of fermion and spin states $|\Psi_0\rangle =  \left|S \right>_{\mathrm{\sigma}} \otimes \left|\psi \right\rangle_\mathrm{f} $ as the initial state. In order to investigate the dynamics of fermions on the dual language, the initial state is transformed to the one in the Hilbert space of the $\hat{C}$ degrees of freedom. If we choose the fermion states as the Fock states, i.e., $\left|\psi \right\rangle_\mathrm{f} = \hat{f}^{\dagger}_i \cdots \hat{f}^{\dagger}_j \left|\mathrm{vacuum} \right\rangle $, then the states have the same form for $\hat{C}$ fermions $\left|\psi \right\rangle_\mathrm{C} = \hat{C}^{\dagger}_i \cdots \hat{C}^{\dagger}_j \left|\mathrm{vacuum} \right\rangle $.  In this article,
the spin states are restricted to the z-polarized state  $|S\rangle_{\mathrm{\sigma}}=|\uparrow \uparrow \uparrow \cdots\rangle$. From the duality transformation, we can obtain that
\begin{eqnarray}
\hat{\tau}_i^x \hat{\tau}_j^x  |\uparrow \uparrow \uparrow \cdots\rangle_{\mathrm{\sigma}} =\prod_{i} \hat{\tau}_i^z |\uparrow \uparrow \uparrow \cdots\rangle_{\mathrm{\sigma}}=|\uparrow \uparrow \uparrow \cdots\rangle_{\mathrm{\sigma}},
\end{eqnarray}
where the relation $|\uparrow \uparrow \uparrow \cdots\rangle_{\mathrm{\sigma}} = \frac{1}{\sqrt{2}} (|\rightarrow  \rightarrow \cdots\rangle_{\mathrm{\tau}}+|\leftarrow \leftarrow \cdots\rangle_{\mathrm{\tau}} )$ has been used.
Then the initial state form $\left|S \right>_{\mathrm{\sigma}} \otimes  \left|\psi \right\rangle_\mathrm{f}  \propto \left|S \right>_{\mathrm{\tau}}
\otimes  \left|\psi \right\rangle_\mathrm{C}  $ can be identified. Next,  the $\tau$ spins  can be replaced by the conserved charges,

\begin{eqnarray}
|\uparrow \uparrow \uparrow \cdots\rangle_{\mathrm{\sigma}} \otimes \left|\psi\right\rangle  = \frac{1}{\sqrt{2^{N-1}}}
\sum_{\left\langle \tau_i \right\rangle= \uparrow \downarrow } \left|\tau_1, \tau_2, \cdots \right\rangle \otimes \left|\psi\right\rangle_\mathrm{C},
\end{eqnarray}
where $\left| \rightarrow \right\rangle_{\mathrm{\tau}} = \left(\left|\uparrow \right\rangle_{\mathrm{\tau}} + \left|\downarrow \right\rangle_{\mathrm{\tau}} \right)$ for each $\tau$ spin and the sum is over all charge configurations. Taking a single state $\left|\tau_1, \tau_2, \cdots \right\rangle \otimes \left|\psi\right\rangle_\mathrm{C}$ as an example, the fermion state is the tensor product of site occupation,  this single state is rewritten as
\begin{eqnarray}
\left|(-1)^{n_1} \tau_1, (-1)^{n_2} \tau_2, \cdots \right\rangle_\mathrm{q} \otimes \left|\psi\right\rangle_\mathrm{C}.
\end{eqnarray}
At the initial state, since the  occupation numbers are fixed, only a common sign structure is contributed. Therefore, the initial state is transformed  into a dual representation, see Eq.~(\ref{initial state}).
In this work, we used QuSpin for simulating the dynamics of the
systems \cite{SciPostPhys.2.1.003,SciPostPhys.7.2.020}.

\section{ Data Availability}
The data that support the findings of this study are available from the corresponding authors upon reasonable request.

\section{ Code availability}
The code used for the analysis is available from the authors upon reasonable request.

\section{ACKNOWLEDGMENTS}
This project is supported by the National Key R$\&$D Program of China, Grants No. 2022YFA1402802, No. 2018YFA0306001, NSFC-92165204, NSFC-11974432, and Shenzhen Institute for Quantum Science and Engineering (Grant No. SIQSE202102). J.Q.C. is supported by the Special Fund of Theoretical Physics of National Nature Science Foundation of China through Grants No.12047562.  S.Y. is supported by the National Natural Science Foundation of China (Grants No. 12222515 and No. 12075324) and the Science and Technology Projects in Guangdong Province (Grants No. 2021QN02X561). 

\section{ AUTHOR CONTRIBUTIONS}
S.Y., D.X.Y., and J.Q.C. conceived and designed the project. J.Q.C. performed the numerical simulations.  J.Q.C., S.Y.,and D.X.Y. provided the explanation of the numerical results. All authors contributed to the discussion of the results and wrote the paper.

\section{ Competing Interests}
The authors declare no competing interests.


	\clearpage


\section{References}
	

\begin{thebibliography}{10}
	\expandafter\ifx\csname url\endcsname\relax
	  \def\url#1{\texttt{#1}}\fi
	\expandafter\ifx\csname urlprefix\endcsname\relax\def\urlprefix{URL }\fi
	\providecommand{\bibinfo}[2]{#2}
	\providecommand{\eprint}[2][]{\url{#2}}
	
	\bibitem{wen2004quantum}
	\bibinfo{author}{Wen, X.-G.}
	\newblock \emph{\bibinfo{title}{Quantum Field Theory of Many-Body Systems}}
	  (\bibinfo{publisher}{Oxford University Press}, \bibinfo{address}{Oxford},
	  \bibinfo{year}{2004}).
	
	\bibitem{Fradkin2013FieldTO}
	\bibinfo{author}{Fradkin, E.}
	\newblock \bibinfo{title}{Field theories of condensed matter physics}
	  (\bibinfo{publisher}{Cambridge University Press},
	  \bibinfo{address}{Cambridge}, \bibinfo{year}{2013}).
	
	\bibitem{PhysRevD.10.2445}
	\bibinfo{author}{Wilson, K.~G.}
	\newblock \bibinfo{title}{Confinement of quarks}.
	\newblock \emph{\bibinfo{journal}{Phys. Rev. D}} \textbf{\bibinfo{volume}{10}},
	  \bibinfo{pages}{2445--2459} (\bibinfo{year}{1974}).
	
	\bibitem{PhysRevLett.62.1694}
	\bibinfo{author}{Read, N.} \& \bibinfo{author}{Sachdev, S.}
	\newblock \bibinfo{title}{Valence-bond and spin-{Peierls} ground states of
	  low-dimensional quantum antiferromagnets}.
	\newblock \emph{\bibinfo{journal}{Phys. Rev. Lett.}}
	  \textbf{\bibinfo{volume}{62}}, \bibinfo{pages}{1694--1697}
	  (\bibinfo{year}{1989}).
	
	\bibitem{PhysRevB.37.580}
	\bibinfo{author}{Baskaran, G.} \& \bibinfo{author}{Anderson, P.~W.}
	\newblock \bibinfo{title}{Gauge theory of high-temperature superconductors and
	  strongly correlated {Fermi} systems}.
	\newblock \emph{\bibinfo{journal}{Phys. Rev. B}} \textbf{\bibinfo{volume}{37}},
	  \bibinfo{pages}{580--583} (\bibinfo{year}{1988}).
	
	\bibitem{martinez2016real}
	\bibinfo{author}{Martinez, E.~A.} \emph{et~al.}
	\newblock \bibinfo{title}{Real-time dynamics of lattice gauge theories with a
	  few-qubit quantum computer}.
	\newblock \emph{\bibinfo{journal}{Nature}} \textbf{\bibinfo{volume}{534}},
	  \bibinfo{pages}{516--519} (\bibinfo{year}{2016}).
	
	\bibitem{PhysRevLett.118.070501}
	\bibinfo{author}{Zohar, E.}, \bibinfo{author}{Farace, A.},
	  \bibinfo{author}{Reznik, B.} \& \bibinfo{author}{Cirac, J.~I.}
	\newblock \bibinfo{title}{Digital quantum simulation of $\mathrm{Z}_2$ lattice
	  gauge theories with dynamical fermionic matter}.
	\newblock \emph{\bibinfo{journal}{Phys. Rev. Lett.}}
	  \textbf{\bibinfo{volume}{118}}, \bibinfo{pages}{070501}
	  (\bibinfo{year}{2017}).
	
	\bibitem{schweizer2019}
	\bibinfo{author}{Schweizer, C.} \emph{et~al.}
	\newblock \bibinfo{title}{Floquet approach to $\mathbb{Z}_2$ lattice gauge
	  theories with ultracold atoms in optical lattices}.
	\newblock \emph{\bibinfo{journal}{Nat. Phys.}} \textbf{\bibinfo{volume}{15}},
	  \bibinfo{pages}{1168–1173} (\bibinfo{year}{2019}).
	
	\bibitem{NP2019realization}
	\bibinfo{author}{G{\"o}rg, F.} \emph{et~al.}
	\newblock \bibinfo{title}{Realization of density-dependent {Peierls} phases to
	  engineer quantized gauge fields coupled to ultracold matter}.
	\newblock \emph{\bibinfo{journal}{Nat. Phys.}} \textbf{\bibinfo{volume}{15}},
	  \bibinfo{pages}{1161--1167} (\bibinfo{year}{2019}).
	
	\bibitem{science.JianweiPan}
	\bibinfo{author}{Zhou, Z.-Y.} \emph{et~al.}
	\newblock \bibinfo{title}{Thermalization dynamics of a gauge theory on a
	  quantum simulator}.
	\newblock \emph{\bibinfo{journal}{Science}} \textbf{\bibinfo{volume}{377}},
	  \bibinfo{pages}{311--314} (\bibinfo{year}{2022}).
	
	\bibitem{pan_2020}
	\bibinfo{author}{Yang, B.} \emph{et~al.}
	\newblock \bibinfo{title}{Observation of gauge invariance in a 71-site
	  {Bose–Hubbard} quantum simulator}.
	\newblock \emph{\bibinfo{journal}{Nature}} \textbf{\bibinfo{volume}{587}},
	  \bibinfo{pages}{392–396} (\bibinfo{year}{2020}).
	
	\bibitem{Banuls2020}
	\bibinfo{author}{Bañuls, M.} \emph{et~al.}
	\newblock \bibinfo{title}{Simulating lattice gauge theories within quantum
	  technologies}.
	\newblock \emph{\bibinfo{journal}{Eur. Phys. J. D}}
	  \textbf{\bibinfo{volume}{74}} (\bibinfo{year}{2020}).
	
	\bibitem{science367}
	\bibinfo{author}{Mil, A.} \emph{et~al.}
	\newblock \bibinfo{title}{A scalable realization of local ${U(1)}$ gauge
	  invariance in cold atomic mixtures}.
	\newblock \emph{\bibinfo{journal}{Science}} \textbf{\bibinfo{volume}{367}},
	  \bibinfo{pages}{1128--1130} (\bibinfo{year}{2020}).
	
	\bibitem{scienceadvance2019}
	\bibinfo{author}{Barbiero, L.} \emph{et~al.}
	\newblock \bibinfo{title}{Coupling ultracold matter to dynamical gauge fields
	  in optical lattices: From flux attachment to $\mathrm{Z}_2$ lattice gauge
	  theories}.
	\newblock \emph{\bibinfo{journal}{Sci. Adv.}} \textbf{\bibinfo{volume}{5}},
	  \bibinfo{pages}{eaav7444} (\bibinfo{year}{2019}).
	
	\bibitem{PhysRevResearch.4.L022060}
	\bibinfo{author}{Wang, Z.} \emph{et~al.}
	\newblock \bibinfo{title}{Observation of emergent $\mathrm{Z}_2$ gauge
	  invariance in a superconducting circuit}.
	\newblock \emph{\bibinfo{journal}{Phys. Rev. Research}}
	  \textbf{\bibinfo{volume}{4}}, \bibinfo{pages}{L022060}
	  (\bibinfo{year}{2022}).
	
	\bibitem{PhysRevX.10.021041}
	\bibinfo{author}{Surace, F.~M.} \emph{et~al.}
	\newblock \bibinfo{title}{Lattice gauge theories and string dynamics in
	  {Rydberg} atom quantum simulators}.
	\newblock \emph{\bibinfo{journal}{Phys. Rev. X}} \textbf{\bibinfo{volume}{10}},
	  \bibinfo{pages}{021041} (\bibinfo{year}{2020}).
	
	\bibitem{PhysRevResearch.4.033120}
	\bibinfo{author}{Halimeh, J.~C.} \emph{et~al.}
	\newblock \bibinfo{title}{Stabilizing lattice gauge theories through simplified
	  local pseudogenerators}.
	\newblock \emph{\bibinfo{journal}{Phys. Rev. Res.}}
	  \textbf{\bibinfo{volume}{4}}, \bibinfo{pages}{033120} (\bibinfo{year}{2022}).
	
	\bibitem{PhysRevLett.118.266601}
	\bibinfo{author}{Smith, A.}, \bibinfo{author}{Knolle, J.},
	  \bibinfo{author}{Kovrizhin, D.~L.} \& \bibinfo{author}{Moessner, R.}
	\newblock \bibinfo{title}{Disorder-free localization}.
	\newblock \emph{\bibinfo{journal}{Phys. Rev. Lett.}}
	  \textbf{\bibinfo{volume}{118}}, \bibinfo{pages}{266601}
	  (\bibinfo{year}{2017}).
	
	\bibitem{thesis}
	\bibinfo{author}{Smith, A.}
	\newblock \emph{\bibinfo{title}{Disorder-Free Localization}}
	  (\bibinfo{publisher}{Springer International Publishing},
	  \bibinfo{address}{New York}, \bibinfo{year}{2019}).
	
	\bibitem{PhysRevLett.120.030601}
	\bibinfo{author}{Brenes, M.}, \bibinfo{author}{Dalmonte, M.},
	  \bibinfo{author}{Heyl, M.} \& \bibinfo{author}{Scardicchio, A.}
	\newblock \bibinfo{title}{Many-body localization dynamics from gauge
	  invariance}.
	\newblock \emph{\bibinfo{journal}{Phys. Rev. Lett.}}
	  \textbf{\bibinfo{volume}{120}}, \bibinfo{pages}{030601}
	  (\bibinfo{year}{2018}).
	
	\bibitem{PhysRevB.102.104302}
	\bibinfo{author}{Yao, Z.}, \bibinfo{author}{Liu, C.}, \bibinfo{author}{Zhang,
	  P.} \& \bibinfo{author}{Zhai, H.}
	\newblock \bibinfo{title}{Many-body localization from dynamical gauge fields}.
	\newblock \emph{\bibinfo{journal}{Phys. Rev. B}}
	  \textbf{\bibinfo{volume}{102}}, \bibinfo{pages}{104302}
	  (\bibinfo{year}{2020}).
	
	\bibitem{PhysRevA.103.022416}
	\bibinfo{author}{Danac\ifmmode \imath \else~\i \fi{}, B.} \emph{et~al.}
	\newblock \bibinfo{title}{Disorder-free localization in quantum walks}.
	\newblock \emph{\bibinfo{journal}{Phys. Rev. A}}
	  \textbf{\bibinfo{volume}{103}}, \bibinfo{pages}{022416}
	  (\bibinfo{year}{2021}).
	
	\bibitem{PRXQuantum.3.020345}
	\bibinfo{author}{Halimeh, J.~C.} \emph{et~al.}
	\newblock \bibinfo{title}{Enhancing disorder-free localization through
	  dynamically emergent local symmetries}.
	\newblock \emph{\bibinfo{journal}{PRX Quantum}} \textbf{\bibinfo{volume}{3}},
	  \bibinfo{pages}{020345} (\bibinfo{year}{2022}).
	
	\bibitem{halimeh2022temperature}
	\bibinfo{author}{Halimeh, J.~C.}, \bibinfo{author}{Hauke, P.},
	  \bibinfo{author}{Knolle, J.} \& \bibinfo{author}{Grusdt, F.}
	\newblock \bibinfo{title}{Temperature-induced disorder-free localization} \newblock \eprint{arXiv:2206.11273}
	  (\bibinfo{year}{2022}).
	
	\bibitem{RevModPhys.88.035002}
	\bibinfo{author}{Konotop, V.~V.}, \bibinfo{author}{Yang, J.} \&
	  \bibinfo{author}{Zezyulin, D.~A.}
	\newblock \bibinfo{title}{Nonlinear waves in $\mathcal{PT}$-symmetric systems}.
	\newblock \emph{\bibinfo{journal}{Rev. Mod. Phys.}}
	  \textbf{\bibinfo{volume}{88}}, \bibinfo{pages}{035002}
	  (\bibinfo{year}{2016}).
	
	\bibitem{ElGanainy2018}
	\bibinfo{author}{El-Ganainy, R.} \emph{et~al.}
	\newblock \bibinfo{title}{Non-hermitian physics and $\mathcal{PT}$ symmetry}.
	\newblock \emph{\bibinfo{journal}{Nat. Phys.}} \textbf{\bibinfo{volume}{14}},
	  \bibinfo{pages}{11--19} (\bibinfo{year}{2018}).
	
	\bibitem{Mirieaar7709}
	\bibinfo{author}{Miri, M.-A.} \& \bibinfo{author}{Al{\`u}, A.}
	\newblock \bibinfo{title}{Exceptional points in optics and photonics}.
	\newblock \emph{\bibinfo{journal}{Science}} \textbf{\bibinfo{volume}{363}}
	  (\bibinfo{year}{2019}).
	
	\bibitem{PhysRevLett.116.133903}
	\bibinfo{author}{Lee, T.~E.}
	\newblock \bibinfo{title}{Anomalous edge state in a non-{Hermitian} lattice}.
	\newblock \emph{\bibinfo{journal}{Phys. Rev. Lett.}}
	  \textbf{\bibinfo{volume}{116}}, \bibinfo{pages}{133903}
	  (\bibinfo{year}{2016}).
	
	\bibitem{PhysRevLett.121.086803}
	\bibinfo{author}{Yao, S.} \& \bibinfo{author}{Wang, Z.}
	\newblock \bibinfo{title}{Edge states and topological invariants of
	  non-{Hermitian} systems}.
	\newblock \emph{\bibinfo{journal}{Phys. Rev. Lett.}}
	  \textbf{\bibinfo{volume}{121}}, \bibinfo{pages}{086803}
	  (\bibinfo{year}{2018}).
	
	\bibitem{PhysRevLett.121.026808}
	\bibinfo{author}{Kunst, F.~K.}, \bibinfo{author}{Edvardsson, E.},
	  \bibinfo{author}{Budich, J.~C.} \& \bibinfo{author}{Bergholtz, E.~J.}
	\newblock \bibinfo{title}{Biorthogonal bulk-boundary correspondence in
	  non-{Hermitian} systems}.
	\newblock \emph{\bibinfo{journal}{Phys. Rev. Lett.}}
	  \textbf{\bibinfo{volume}{121}}, \bibinfo{pages}{026808}
	  (\bibinfo{year}{2018}).
	
	\bibitem{PhysRevX.9.041015}
	\bibinfo{author}{Kawabata, K.}, \bibinfo{author}{Shiozaki, K.},
	  \bibinfo{author}{Ueda, M.} \& \bibinfo{author}{Sato, M.}
	\newblock \bibinfo{title}{Symmetry and topology in non-{Hermitian} physics}.
	\newblock \emph{\bibinfo{journal}{Phys. Rev. X}} \textbf{\bibinfo{volume}{9}},
	  \bibinfo{pages}{041015} (\bibinfo{year}{2019}).
	
	\bibitem{PhysRevLett.123.090603}
	\bibinfo{author}{Hamazaki, R.}, \bibinfo{author}{Kawabata, K.} \&
	  \bibinfo{author}{Ueda, M.}
	\newblock \bibinfo{title}{Non-{Hermitian} many-body localization}.
	\newblock \emph{\bibinfo{journal}{Phys. Rev. Lett.}}
	  \textbf{\bibinfo{volume}{123}}, \bibinfo{pages}{090603}
	  (\bibinfo{year}{2019}).
	
	\bibitem{PhysRevB.100.054301}
	\bibinfo{author}{Jiang, H.}, \bibinfo{author}{Lang, L.-J.},
	  \bibinfo{author}{Yang, C.}, \bibinfo{author}{Zhu, S.-L.} \&
	  \bibinfo{author}{Chen, S.}
	\newblock \bibinfo{title}{Interplay of non-{Hermitian} skin effects and
	  anderson localization in nonreciprocal quasiperiodic lattices}.
	\newblock \emph{\bibinfo{journal}{Phys. Rev. B}}
	  \textbf{\bibinfo{volume}{100}}, \bibinfo{pages}{054301}
	  (\bibinfo{year}{2019}).
	
	\bibitem{PhysRevB.102.064206}
	\bibinfo{author}{Zhai, L.-J.}, \bibinfo{author}{Yin, S.} \&
	  \bibinfo{author}{Huang, G.-Y.}
	\newblock \bibinfo{title}{Many-body localization in a {non-Hermitian}
	  quasiperiodic system}.
	\newblock \emph{\bibinfo{journal}{Phys. Rev. B}}
	  \textbf{\bibinfo{volume}{102}}, \bibinfo{pages}{064206}
	  (\bibinfo{year}{2020}).
	
	\bibitem{PhysRevLett.126.166801}
	\bibinfo{author}{Kawabata, K.} \& \bibinfo{author}{Ryu, S.}
	\newblock \bibinfo{title}{Nonunitary scaling theory of non-{Hermitian}
	  localization}.
	\newblock \emph{\bibinfo{journal}{Phys. Rev. Lett.}}
	  \textbf{\bibinfo{volume}{126}}, \bibinfo{pages}{166801}
	  (\bibinfo{year}{2021}).
	
	\bibitem{Naturephysics2010}
	\bibinfo{author}{R{\"u}ter, C.~E.} \emph{et~al.}
	\newblock \bibinfo{title}{Observation of parity--time symmetry in optics}.
	\newblock \emph{\bibinfo{journal}{Nat. Phys.}} \textbf{\bibinfo{volume}{6}},
	  \bibinfo{pages}{192--195} (\bibinfo{year}{2010}).
	
	\bibitem{Scienceadv2016}
	\bibinfo{author}{Ma, G.} \& \bibinfo{author}{Sheng, P.}
	\newblock \bibinfo{title}{Acoustic metamaterials: From local resonances to
	  broad horizons}.
	\newblock \emph{\bibinfo{journal}{Sci. Adv.}} \textbf{\bibinfo{volume}{2}},
	  \bibinfo{pages}{e1501595} (\bibinfo{year}{2016}).
	
	\bibitem{NC2019observation}
	\bibinfo{author}{Li, J.} \emph{et~al.}
	\newblock \bibinfo{title}{Observation of parity-time symmetry breaking
	  transitions in a dissipative floquet system of ultracold atoms}.
	\newblock \emph{\bibinfo{journal}{Nat. Commun.}} \textbf{\bibinfo{volume}{10}},
	  \bibinfo{pages}{855} (\bibinfo{year}{2019}).
	
	\bibitem{aidelsburger2022cold}
	\bibinfo{author}{Aidelsburger, M.} \emph{et~al.}
	\newblock \bibinfo{title}{Cold atoms meet lattice gauge theory}.
	\newblock \emph{\bibinfo{journal}{Philos. T. R. Soc. A}}
	  \textbf{\bibinfo{volume}{380}}, \bibinfo{pages}{20210064}
	  (\bibinfo{year}{2022}).
	
	\bibitem{PhysRevLett.77.570}
	\bibinfo{author}{Hatano, N.} \& \bibinfo{author}{Nelson, D.~R.}
	\newblock \bibinfo{title}{Localization transitions in non-{Hermitian} quantum
	  mechanics}.
	\newblock \emph{\bibinfo{journal}{Phys. Rev. Lett.}}
	  \textbf{\bibinfo{volume}{77}}, \bibinfo{pages}{570--573}
	  (\bibinfo{year}{1996}).
	
	\bibitem{PhysRevLett.80.2897}
	\bibinfo{author}{Goldsheid, I.~Y.} \& \bibinfo{author}{Khoruzhenko, B.~A.}
	\newblock \bibinfo{title}{Distribution of eigenvalues in non-{Hermitian}
	  {Anderson} models}.
	\newblock \emph{\bibinfo{journal}{Phys. Rev. Lett.}}
	  \textbf{\bibinfo{volume}{80}}, \bibinfo{pages}{2897--2900}
	  (\bibinfo{year}{1998}).
	
	\bibitem{PhysRevB.103.054203}
	\bibinfo{author}{Longhi, S.}
	\newblock \bibinfo{title}{Phase transitions in a non-{Hermitian}
	  {Aubry-Andr\'e-Harper} model}.
	\newblock \emph{\bibinfo{journal}{Phys. Rev. B}}
	  \textbf{\bibinfo{volume}{103}}, \bibinfo{pages}{054203}
	  (\bibinfo{year}{2021}).
	
	\bibitem{PhysRevA.95.062118}
	\bibinfo{author}{Zeng, Q.-B.}, \bibinfo{author}{Chen, S.} \&
	  \bibinfo{author}{L\"u, R.}
	\newblock \bibinfo{title}{Anderson localization in the non-{Hermitian}
	  {Aubry}-{Andr\'e}-{Harper} model with physical gain and loss}.
	\newblock \emph{\bibinfo{journal}{Phys. Rev. A}} \textbf{\bibinfo{volume}{95}},
	  \bibinfo{pages}{062118} (\bibinfo{year}{2017}).
	
	\bibitem{PhysRevB.101.174205}
	\bibinfo{author}{Liu, Y.}, \bibinfo{author}{Jiang, X.-P.},
	  \bibinfo{author}{Cao, J.} \& \bibinfo{author}{Chen, S.}
	\newblock \bibinfo{title}{Non-{Hermitian} mobility edges in one-dimensional
	  quasicrystals with parity-time symmetry}.
	\newblock \emph{\bibinfo{journal}{Phys. Rev. B}}
	  \textbf{\bibinfo{volume}{101}}, \bibinfo{pages}{174205}
	  (\bibinfo{year}{2020}).
	
	\bibitem{PhysRevB.106.014204}
	\bibinfo{author}{Zhai, L.-J.}, \bibinfo{author}{Huang, G.-Y.} \&
	  \bibinfo{author}{Yin, S.}
	\newblock \bibinfo{title}{Nonequilibrium dynamics of the
	  localization-delocalization transition in the {non-Hermitian} {Aubry-Andr\'e}
	  model}.
	\newblock \emph{\bibinfo{journal}{Phys. Rev. B}}
	  \textbf{\bibinfo{volume}{106}}, \bibinfo{pages}{014204}
	  (\bibinfo{year}{2022}).
	
	\bibitem{PhysRevD.17.2637}
	\bibinfo{author}{Fradkin, E.} \& \bibinfo{author}{Susskind, L.}
	\newblock \bibinfo{title}{Order and disorder in gauge systems and magnets}.
	\newblock \emph{\bibinfo{journal}{Phys. Rev. D}} \textbf{\bibinfo{volume}{17}},
	  \bibinfo{pages}{2637--2658} (\bibinfo{year}{1978}).
	
	\bibitem{PhysRevB.96.205104}
	\bibinfo{author}{Prosko, C.}, \bibinfo{author}{Lee, S.-P.} \&
	  \bibinfo{author}{Maciejko, J.}
	\newblock \bibinfo{title}{Simple ${\mathbb{z}}_{2}$ lattice gauge theories at
	  finite fermion density}.
	\newblock \emph{\bibinfo{journal}{Phys. Rev. B}} \textbf{\bibinfo{volume}{96}},
	  \bibinfo{pages}{205104} (\bibinfo{year}{2017}).
	
	\bibitem{S2020topological}
	\bibinfo{author}{Weidemann, S.} \emph{et~al.}
	\newblock \bibinfo{title}{Topological funneling of light}.
	\newblock \emph{\bibinfo{journal}{Science}} \textbf{\bibinfo{volume}{368}},
	  \bibinfo{pages}{311--314} (\bibinfo{year}{2020}).
	
	\bibitem{S2021generating}
	\bibinfo{author}{Wang, K.} \emph{et~al.}
	\newblock \bibinfo{title}{Generating arbitrary topological windings of a
	  non-Hermitian band}.
	\newblock \emph{\bibinfo{journal}{Science}} \textbf{\bibinfo{volume}{371}},
	  \bibinfo{pages}{1240--1245} (\bibinfo{year}{2021}).
	
	\bibitem{NP2020generalized}
	\bibinfo{author}{Helbig, T.} \emph{et~al.}
	\newblock \bibinfo{title}{Generalized bulk--boundary correspondence in
	  non-{Hermitian} topolectrical circuits}.
	\newblock \emph{\bibinfo{journal}{Nat. Phys.}} \textbf{\bibinfo{volume}{16}},
	  \bibinfo{pages}{747--750} (\bibinfo{year}{2020}).
	
	\bibitem{PhysRevX.8.031079}
	\bibinfo{author}{Gong, Z.} \emph{et~al.}
	\newblock \bibinfo{title}{Topological phases of non-{Hermitian} systems}.
	\newblock \emph{\bibinfo{journal}{Phys. Rev. X}} \textbf{\bibinfo{volume}{8}},
	  \bibinfo{pages}{031079} (\bibinfo{year}{2018}).
	
	\bibitem{PhysRevLett.97.050401}
	\bibinfo{author}{Bravyi, S.}, \bibinfo{author}{Hastings, M.~B.} \&
	  \bibinfo{author}{Verstraete, F.}
	\newblock \bibinfo{title}{Lieb-robinson bounds and the generation of
	  correlations and topological quantum order}.
	\newblock \emph{\bibinfo{journal}{Phys. Rev. Lett.}}
	  \textbf{\bibinfo{volume}{97}}, \bibinfo{pages}{050401}
	  (\bibinfo{year}{2006}).
	
	\bibitem{Essler_2016}
	\bibinfo{author}{Essler, F. H.~L.} \& \bibinfo{author}{Fagotti, M.}
	\newblock \bibinfo{title}{Quench dynamics and relaxation in isolated integrable
	  quantum spin chains}.
	\newblock \emph{\bibinfo{journal}{J. Stat. Mech.-Theory E.}}
	  \textbf{\bibinfo{volume}{2016}}, \bibinfo{pages}{064002}
	  (\bibinfo{year}{2016}).
	
	\bibitem{PhysRevB.105.024303}
	\bibinfo{author}{Orito, T.} \& \bibinfo{author}{Imura, K.-I.}
	\newblock \bibinfo{title}{Unusual wave-packet spreading and entanglement
	  dynamics in non-{Hermitian} disordered many-body systems}.
	\newblock \emph{\bibinfo{journal}{Phys. Rev. B}}
	  \textbf{\bibinfo{volume}{105}}, \bibinfo{pages}{024303}
	  (\bibinfo{year}{2022}).
	
	\bibitem{PhysRevB.56.R4333}
	\bibinfo{author}{Brouwer, P.~W.}, \bibinfo{author}{Silvestrov, P.~G.} \&
	  \bibinfo{author}{Beenakker, C. W.~J.}
	\newblock \bibinfo{title}{Theory of directed localization in one dimension}.
	\newblock \emph{\bibinfo{journal}{Phys. Rev. B}} \textbf{\bibinfo{volume}{56}},
	  \bibinfo{pages}{R4333--R4335} (\bibinfo{year}{1997}).
	
	\bibitem{PhysRevB.56.8651}
	\bibinfo{author}{Hatano, N.} \& \bibinfo{author}{Nelson, D.~R.}
	\newblock \bibinfo{title}{Vortex pinning and non-{Hermitian} quantum
	  mechanics}.
	\newblock \emph{\bibinfo{journal}{Phys. Rev. B}} \textbf{\bibinfo{volume}{56}},
	  \bibinfo{pages}{8651--8673} (\bibinfo{year}{1997}).
	
	\bibitem{PhysRevB.103.064201}
	\bibinfo{author}{Heu\ss{}en, S.}, \bibinfo{author}{White, C.~D.} \&
	  \bibinfo{author}{Refael, G.}
	\newblock \bibinfo{title}{Extracting many-body localization lengths with an
	  imaginary vector potential}.
	\newblock \emph{\bibinfo{journal}{Phys. Rev. B}}
	  \textbf{\bibinfo{volume}{103}}, \bibinfo{pages}{064201}
	  (\bibinfo{year}{2021}).
	
	\bibitem{PhysRevLett.75.117}
	\bibinfo{author}{del Rio, R.}, \bibinfo{author}{Jitomirskaya, S.},
	  \bibinfo{author}{Last, Y.} \& \bibinfo{author}{Simon, B.}
	\newblock \bibinfo{title}{What is localization?}
	\newblock \emph{\bibinfo{journal}{Phys. Rev. Lett.}}
	  \textbf{\bibinfo{volume}{75}}, \bibinfo{pages}{117--119}
	  (\bibinfo{year}{1995}).
	
	\bibitem{BKramer_1993}
	\bibinfo{author}{Kramer, B.} \& \bibinfo{author}{MacKinnon, A.}
	\newblock \bibinfo{title}{Localization: theory and experiment}.
	\newblock \emph{\bibinfo{journal}{Rep. Prog. Phys.}}
	  \textbf{\bibinfo{volume}{56}}, \bibinfo{pages}{1469} (\bibinfo{year}{1993}).
	
	\bibitem{PhysRevB.104.104203}
	\bibinfo{author}{Luo, X.}, \bibinfo{author}{Ohtsuki, T.} \&
	  \bibinfo{author}{Shindou, R.}
	\newblock \bibinfo{title}{Transfer matrix study of the anderson transition in
	  non-hermitian systems}.
	\newblock \emph{\bibinfo{journal}{Phys. Rev. B}}
	  \textbf{\bibinfo{volume}{104}}, \bibinfo{pages}{104203}
	  (\bibinfo{year}{2021}).
	
	\bibitem{Grover_2014}
	\bibinfo{author}{Grover, T.} \& \bibinfo{author}{Fisher, M. P.~A.}
	\newblock \bibinfo{title}{Quantum disentangled liquids}.
	\newblock \emph{\bibinfo{journal}{J. Stat. Mech.-Theory E.}}
	  \textbf{\bibinfo{volume}{2014}}, \bibinfo{pages}{P10010}
	  (\bibinfo{year}{2014}).
	
	\bibitem{PhysRevLett.119.176601}
	\bibinfo{author}{Smith, A.}, \bibinfo{author}{Knolle, J.},
	  \bibinfo{author}{Moessner, R.} \& \bibinfo{author}{Kovrizhin, D.~L.}
	\newblock \bibinfo{title}{Absence of ergodicity without quenched disorder: From
	  quantum disentangled liquids to many-body localization}.
	\newblock \emph{\bibinfo{journal}{Phys. Rev. Lett.}}
	  \textbf{\bibinfo{volume}{119}}, \bibinfo{pages}{176601}
	  (\bibinfo{year}{2017}).
	
	\bibitem{PhysRevX.13.021007}
	\bibinfo{author}{Kawabata, K.}, \bibinfo{author}{Numasawa, T.} \&
	  \bibinfo{author}{Ryu, S.}
	\newblock \bibinfo{title}{Entanglement phase transition induced by the
	  non-{Hermitian} skin effect}.
	\newblock \emph{\bibinfo{journal}{Phys. Rev. X}} \textbf{\bibinfo{volume}{13}},
	  \bibinfo{pages}{021007} (\bibinfo{year}{2023}).
	
	\bibitem{NP2023verification}
	\bibinfo{author}{Tajik, M.} \emph{et~al.}
	\newblock \bibinfo{title}{Verification of the area law of mutual information in
	  a quantum field simulator}.
	\newblock \emph{\bibinfo{journal}{Nat. Phys.}} \bibinfo{pages}{1--5}
	  (\bibinfo{year}{2023}).
	
	\bibitem{dogra2021quantum}
	\bibinfo{author}{Dogra, S.}, \bibinfo{author}{Melnikov, A.~A.} \&
	  \bibinfo{author}{Paraoanu, G.~S.}
	\newblock \bibinfo{title}{Quantum simulation of parity--time symmetry breaking
	  with a superconducting quantum processor}.
	\newblock \emph{\bibinfo{journal}{Commun. Phys.}} \textbf{\bibinfo{volume}{4}},
	  \bibinfo{pages}{26} (\bibinfo{year}{2021}).
	
	\bibitem{PhysRevB.105.205141}
	\bibinfo{author}{Riechert, H.} \emph{et~al.}
	\newblock \bibinfo{title}{Engineering a ${U(1)}$ lattice gauge theory in
	  classical electric circuits}.
	\newblock \emph{\bibinfo{journal}{Phys. Rev. B}}
	  \textbf{\bibinfo{volume}{105}}, \bibinfo{pages}{205141}
	  (\bibinfo{year}{2022}).
	
	\bibitem{ezawa2019non}
	\bibinfo{author}{Ezawa, M.}
	\newblock \bibinfo{title}{Non-hermitian higher-order topological states in
	  nonreciprocal and reciprocal systems with their electric-circuit
	  realization}.
	\newblock \emph{\bibinfo{journal}{Phys. Rev. B}} \textbf{\bibinfo{volume}{99}},
	  \bibinfo{pages}{201411} (\bibinfo{year}{2019}).
	
	\bibitem{Research2021}
	\bibinfo{author}{Liu, S.} \emph{et~al.}
	\newblock \bibinfo{title}{Non-hermitian skin effect in a non-hermitian
	  electrical circuit}.
	\newblock \emph{\bibinfo{journal}{Research}} \textbf{\bibinfo{volume}{2021}},
	  \bibinfo{pages}{5608038} (\bibinfo{year}{2021}).
	
	\bibitem{li2023manybody}
	\bibinfo{author}{Li, H.}, \bibinfo{author}{Wu, H.}, \bibinfo{author}{Zheng, W.}
	  \& \bibinfo{author}{Yi, W.}
	\newblock \bibinfo{title}{Many-body non-hermitian skin effect under dynamic
	  gauge coupling} \newblock \eprint{arXiv:2305.03891} (\bibinfo{year}{2023}).
	
	\bibitem{N2015measuring}
	\bibinfo{author}{Islam, R.} \emph{et~al.}
	\newblock \bibinfo{title}{Measuring entanglement entropy in a quantum many-body
	  system}.
	\newblock \emph{\bibinfo{journal}{Nature}} \textbf{\bibinfo{volume}{528}},
	  \bibinfo{pages}{77--83} (\bibinfo{year}{2015}).
	
	\bibitem{Lukin256}
	\bibinfo{author}{Lukin, A.} \emph{et~al.}
	\newblock \bibinfo{title}{Probing entanglement in a many-body-localized
	  system}.
	\newblock \emph{\bibinfo{journal}{Science}} \textbf{\bibinfo{volume}{364}},
	  \bibinfo{pages}{256--260} (\bibinfo{year}{2019}).
	
	\bibitem{SciPostPhys.2.1.003}
	\bibinfo{author}{Weinberg, P.} \& \bibinfo{author}{Bukov, M.}
	\newblock \bibinfo{title}{{QuSpin: a Python package for dynamics and exact
	  diagonalisation of quantum many body systems part I: spin chains}}.
	\newblock \emph{\bibinfo{journal}{SciPost Phys.}} \textbf{\bibinfo{volume}{2}},
	  \bibinfo{pages}{003} (\bibinfo{year}{2017}).
	
	\bibitem{SciPostPhys.7.2.020}
	\bibinfo{author}{Weinberg, P.} \& \bibinfo{author}{Bukov, M.}
	\newblock \bibinfo{title}{{QuSpin: a Python package for dynamics and exact
	  diagonalisation of quantum many body systems. Part II: bosons, fermions and
	  higher spins}}.
	\newblock \emph{\bibinfo{journal}{SciPost Phys.}} \textbf{\bibinfo{volume}{7}},
	  \bibinfo{pages}{020} (\bibinfo{year}{2019}).
	
	\end{thebibliography}

\renewcommand{\figurename}{Supplementary Figure}
\newcommand{\datta}[1]{\textcolor{red}{#1}}
\allowdisplaybreaks

\renewcommand{\bibsection}{}
\renewcommand{\bibnumfmt}[1]{#1.}


\title{Supplementary information for Dynamical localization transition  in the  non-Hermitian lattice gauge theory}

\author{Jun-Qing Cheng}
\affiliation{State Key Laboratory of Optoelectronic Materials and Technologies, Guangdong Provincial Key Laboratory of Magnetoelectric Physics and Devices, School of Physics, Sun Yat-Sen University, Guangzhou 510275, China}
\affiliation{School of Physical Sciences, Great Bay University, Dongguan 523000, China, and Great Bay Institute for Advanced Study, Dongguan 523000, China}

\author{Shuai Yin}
\email{yinsh6@mail.sysu.edu.cn}
\affiliation{State Key Laboratory of Optoelectronic Materials and Technologies, Guangdong Provincial Key Laboratory of Magnetoelectric Physics and Devices, School of Physics, Sun Yat-Sen University, Guangzhou 510275, China}
\date{\today}

\author{Dao-Xin Yao}
\email{yaodaox@mail.sysu.edu.cn}
\affiliation{State Key Laboratory of Optoelectronic Materials and Technologies, Guangdong Provincial Key Laboratory of Magnetoelectric Physics and Devices, School of Physics, Sun Yat-Sen University, Guangzhou 510275, China}

\maketitle

\begin{center}
  {\centering \bf Supplementary Note 1: Quench dynamics from the domain wall initial states}
  \end{center}

\begin{figure*}[htbp]
	\centering
	\includegraphics[width=16cm]{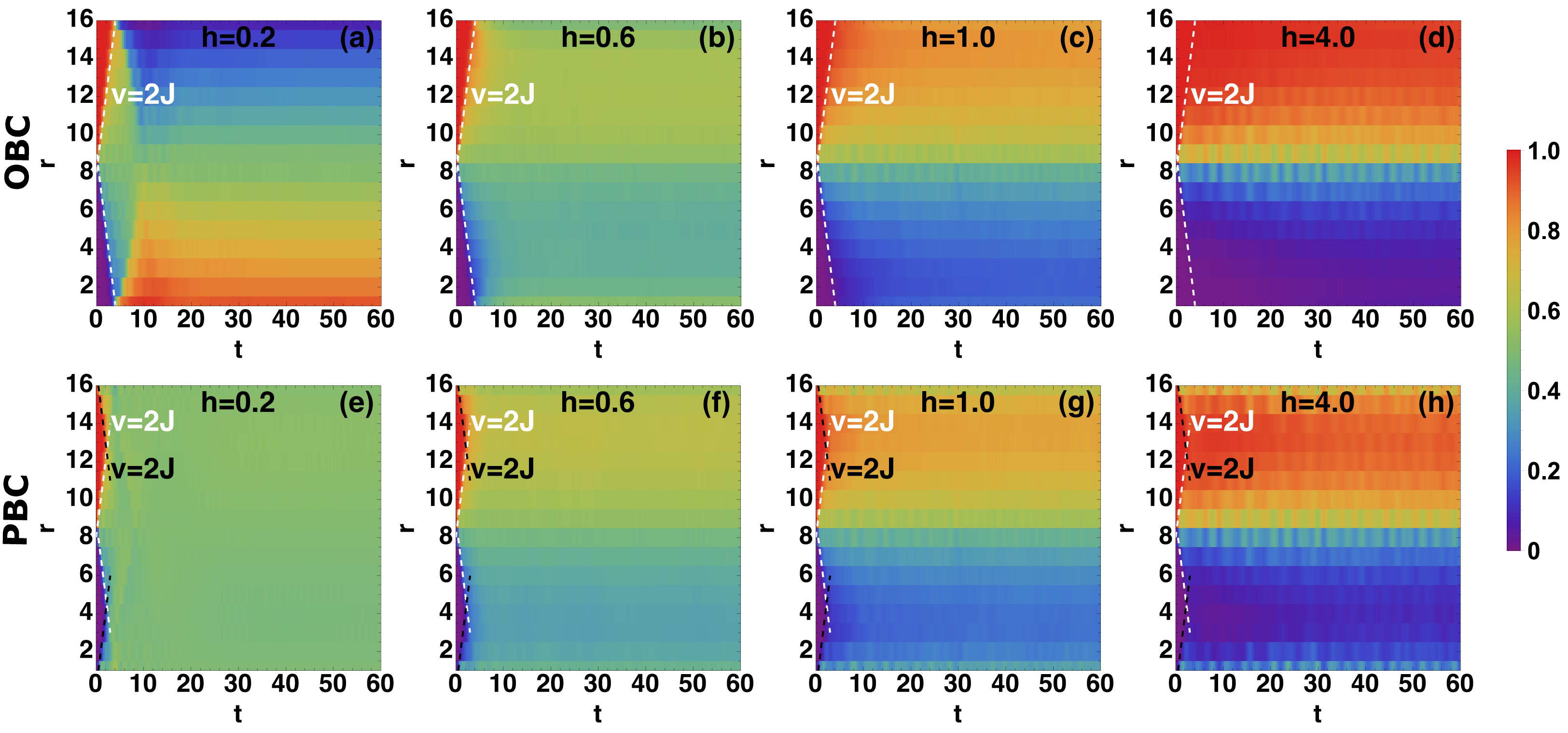}
	\caption{\label{OPBC-DW}  \textbf{Dynamics of fermion density.} Dynamics of fermion density  $\left\langle n_i (t) \right\rangle$  for the system with length $L=16$ under (a)(b)(c)(d) open boundary condition (OBC) and (e)(f)(g)(h) periodic boundary condition (PBC) at different $h$. The initial state of fermionic subsystem is   domain wall (DW) state  and the non-Hermiticity is chosen as $g=0.2$. The white and black dashed lines represent the  propagating signal, half of  Lieb-Robinson velocity $v=v_{\mathrm{LR}}/2=2J$. The color bar indicates the values of fermion density.}
\end{figure*} 

In this section, we  consider the quench dynamics from the domain wall (DW) initial state $\left|0,\dots,0,1,\dots,1 \right\rangle$. In Supplementary Figure ~\ref{OPBC-DW}, the  dynamics of fermion density  $\left\langle n(t) \right\rangle$  for the system with $L=16$ under OBC (see Supplementary Figures ~\ref{OPBC-DW} (a)(b)(c)(d)) and PBC (see Supplementary Figures ~\ref{OPBC-DW} (e)(f)(g)(h)) are displayed. For the OBC case at $h=0.2$, the non-Hermitian skin effect drives the fermions to propagate to the one side, meanwhile the holes transport to in opposite direction. Both the fermions and holes travel at the maximal fermion group velocity $v=2J$ which is also the one half of Lieb-Robinson velocity $v=v_{\mathrm{LR}}/2=2J$ \cite{Essler_2016,thesis}. The weak effective  disorder $h=0.2$ is not strong to preserve the  memory of initial DW state, and an approximate reverse DW state is presents  due to the Pauli exclusion principle.
 As $h$ is increased, the localization behavior becomes more obvious and the memories of initial state are preserved.  At the boundary of domain wall, large $h$ also induces the distinct oscillation in the fermion density and leads the fermions or holes across the domain wall. For the PBC case, the non-Hermitian skin effect disappears. There are two arches at the early time evolution since there are  two domain walls in the initial state. The propagating velocities are still the maximal fermion group velocity $v=2J$. For the weak disorder, the fermions propagate throughout the system, and the fermion subsystem is thermalized.
The localization behavior becomes  more and more clearly as the $h$ is increased.  For strong disorder $h=4.0$, some fermions and holes also pass through the two domain walls because of the oscillations.

\begin{center}
  {\centering \bf Supplementary Note 2: Dynamics of spin subsystem}
  \end{center}

  \begin{figure*}[htbp]
    \centering
    \includegraphics[width=13cm]{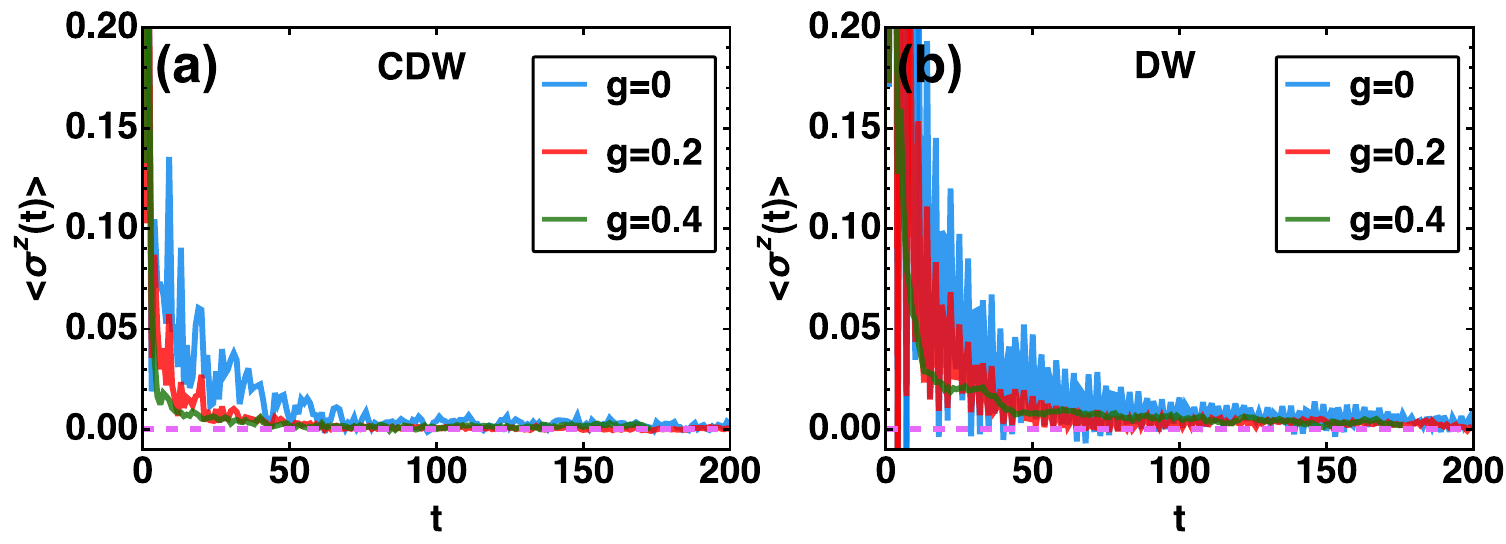}
    \caption{ \label{spinz} \textbf{Dynamics of spin subsystem.} The  time evolution of spin average $\left\langle  \hat{\sigma}^z (t)  \right\rangle$   for different $g$ after the quench from the (a) charge density wave (CDW) state, (b) domain wall (DW) state. All dynamics are measured   under the same condition $L=16$ and $h=1$.}
  \end{figure*}

Above results mainly consider the dynamics of fermion subsystem, now let us  turn to the spin subsystem. The spin correlators can be written as fermionic ones according to the evolution governed by the effective Hamiltonian (see Eq.(2) in the main text). For a polarized initial spin state, the disorder averaged  over all charge configurations  should be considered.
An expectation value of $\hat{\sigma}^z$ spin component on the bond at time $t$ after quench is given by
\begin{eqnarray}
	&& \left\langle  \hat{\sigma}^z_{j,j+1} (t)  \right\rangle =
	\left\langle 0 \right| \mathrm{e}^{\mathrm{i} \hat{H} t}  \hat{\tau}^x_j \hat{\tau}^x_{j+1} \mathrm{e}^{-\mathrm{i} \hat{H} t} \left|0 \right\rangle  \nonumber\\
	&=& \frac{1}{2^{L-1}} \sum_{\left\lbrace q_i \right\rbrace =\pm 1} \left\langle \Psi \right| \mathrm{e}^{\mathrm{i} \hat{H}(q) t}  \mathrm{e}^{-\mathrm{i} \hat{H} (\overline{q}_{j,j+1})  t} \left|\Psi \right\rangle
\end{eqnarray}
where $\overline{q}_{j,j+1}$ means the signs of the charges at $j$ and $j+1$ opposite to the original ones.  The dynamics of spin average is displayed for the charge density wave (CDW) (see  Supplementary Figure ~\ref{spinz} (a)) and domain wall (DW) (see  In Supplementary Figure ~\ref{spinz}(b)) initial states. For the Hermitian case, the magnetizations  both decay to zero with an asymptotic power law. The spin subsystem reach its nonequilibrium states for a short time evolution  along with the  fluctuation, especially for the DW initial state, the fluctuation is more obvious. 
In contrast, the non-Hermiticity effectively suppress the fluctuation and result in the faster decay progresses no matter for the CDW or DW initial states. Although the non-Hermiticity originates from the fermion subsystem, we can still observe its effect on the  dynamics of spin subsystem.

\begin{center}
  {\centering \bf Supplementary Note 3: Entanglement dynamics of fermions}
  \end{center}

The dynamics of entanglement entropy for an isolated system is a well-defined signature for differentiating between thermalization and  localization. Here we focus on the evolution of the half-chain entanglement entropy for the effective system initialized in the CDW state.
  In supplementary Figure \ref{entanglement-C}(a),  except the Hermitian case $g=0$, others $g\geq 0.2$ are in the delocalized phase. For $g=0$, the entanglement entropy grows   after quench and eventually saturates to a  value. The main reason is that  the wave functions initially spreads symmetrically  in the two directions along the system, the entanglement entropy propagates throughout the whole system. After introducing the non-Hermiticity, the entanglement entropy  initially still increases,  but the steady values decrease with the increase of $g$. The non-Hermitian skin effect leads to the antisymmetry spreading of fermionic wave functions   and suppress the spreading process, which contributes to the   nonequilibrium steady state with low entanglement entropy.
Supplementary Figure \ref{entanglement-C}(b) shows the effect of disorder strength on the dynamics of fermions in the localized phase, it can be observed that the entanglement entropy also grows  at the initial period, and the strong disorder  reduces the entanglement entropy of nonequilibrium steady state in the localized phase because of that the wave functions are localized, and the entanglement propagations are suppressed.  In the delocalized and localized phases, the entanglement entropies are both suppressed, but with different reason, the former is due to the non-Hermitian skin effect, while the latter is the consequence of strong effective disorder form the $\mathbb{Z}_2$ gauge field.

\begin{figure*}[htbp]
	\centering
 	\includegraphics[width=12cm]{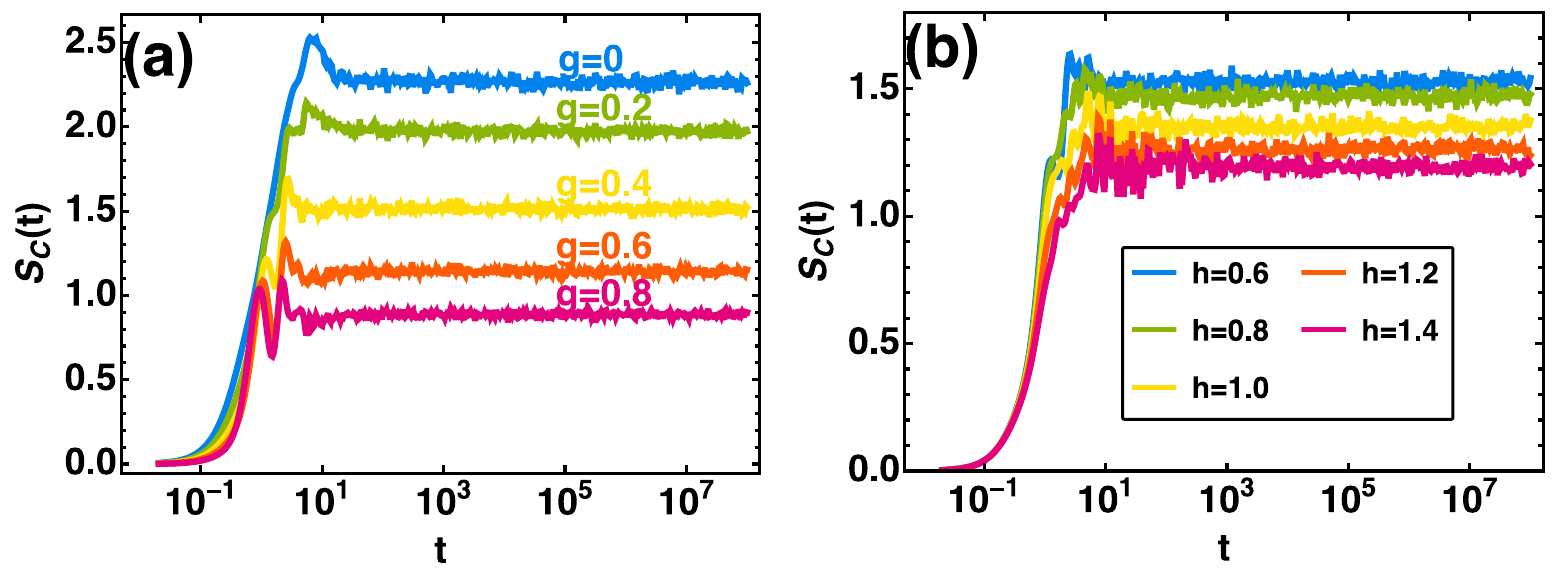}
	\caption{\label{entanglement-C} \textbf{Dynamics of entanglement entropy.} Dynamics of entanglement entropy of fermion  for the open boundary condition with $L=10$ (a) at different hopping strength $g$ for $h=0.5$ and (b) at different effective disorder strength  $h$ for $g=0.4$. The initial state of fermionic subsystem is charge density wave state.  }
\end{figure*}

\begin{figure*}[htbp]
	\centering
	\includegraphics[width=15cm]{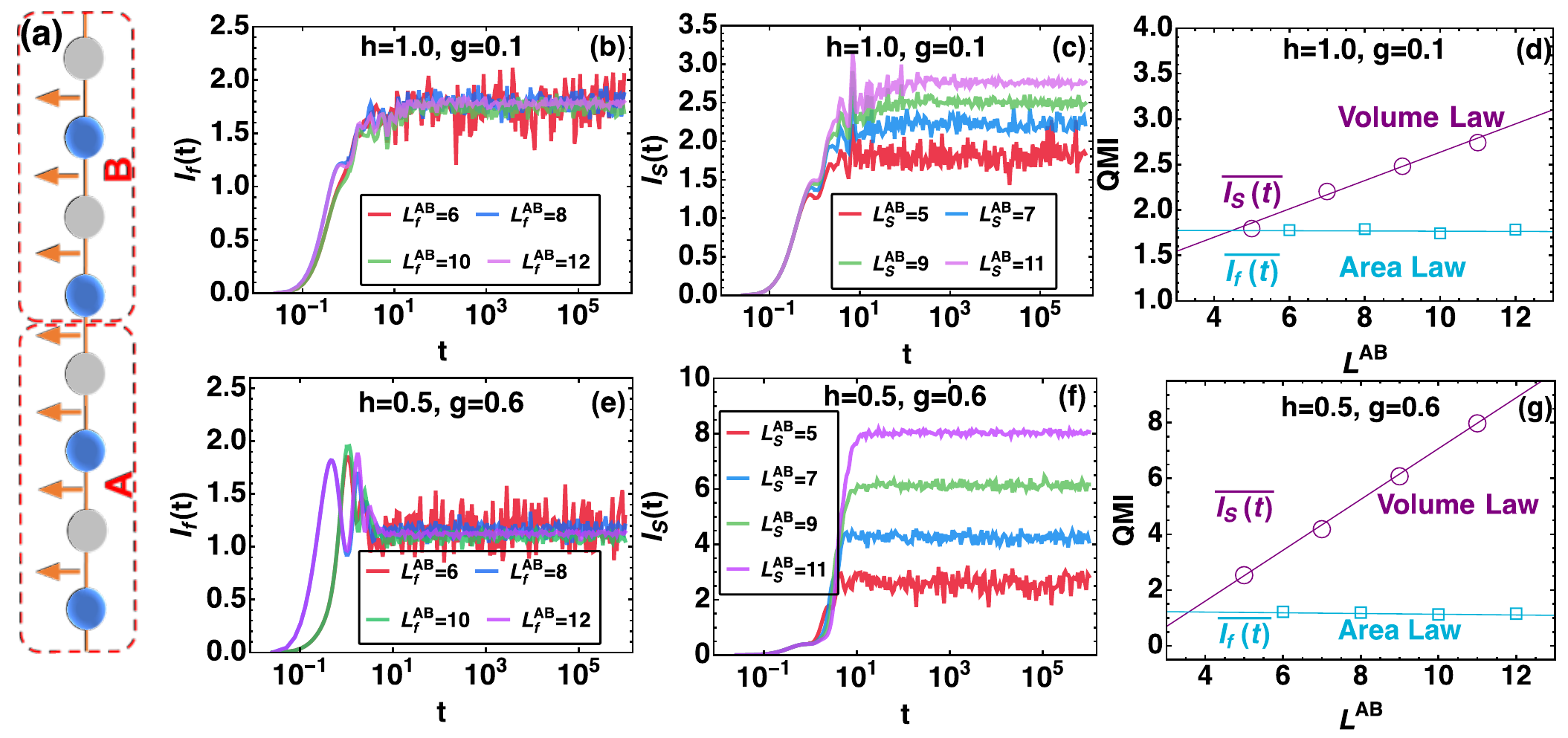}
	\caption{\label{mutual-scaling} \textbf{Features of non-Hermitian quantum disentangled liquid.} (a) The partitioning methods of the whole system.  Dynamics of quantum mutual information  for the fermion and spin subsystems with different sizes at the localized phase (b)(c)(d) $h=1.0, g=0.1 $ and delocalized phase (e)(f)(g) $h=0.5, g=0.6$. The initial state  is charge density wave state for fermion and  the $z$-polarized  state for spins. Saturation values  $ \overline{I_\mathrm{f} (t)}$ and $ \overline{I_\mathrm{S} (t)}$ are averaged from $t=10^3$ to $t=10^6$ for different systems sizes. Here the open boundary condition is considered. }                             
\end{figure*}

\begin{center}
  {\centering \bf Supplementary Note 4: Transfer matrix }
  \end{center}

In this section, we will introduce how to calculate the localization length by transfer matrix method \cite{BKramer_1993,PhysRevLett.126.166801,PhysRevB.104.104203}. This numerical method allows us to compute the Lyapunov exponents for  eigenstates in the thermodynamic limit. The localization length is directly related to the Lyapunov exponents and quantify  the characteristic length for the decay of localized wave functions. We consider the one-dimensional Hamiltonian with onsite binary disorder described by
\begin{eqnarray}
	\hat{H}_{\left\lbrace q_i\right\rbrace} && =-J\sum_i (\mathrm{e}^g \hat{C}_i^{\dagger} \hat{C}_{i+1} +\mathrm{e}^{-g} \hat{C}_{i+1}^{\dagger}\hat{C}_{i}) \nonumber \\
	&&+2h\sum_i q_i \hat{C}_i^{\dagger}\hat{C}_{i},
\end{eqnarray}
where we have omitted an overall energy shift $h\sum_i q_i $ which does not affect the localization length since there are no matrix elements between different charge sectors. An arbitrary single-particle state is given by
\begin{eqnarray}
|\Psi\rangle=\sum_{n=0}^N \Psi_n \hat{C}^{\dagger}|\Omega\rangle,
\end{eqnarray}
where $|\Omega\rangle$ is the fermionic vacuum state and $\Psi_n$ is the particle wave function, the probability amplitude to find the particle at site $n$. The single-particle Schr\"{o}dinger equation $H|\Psi\rangle=E|\Psi\rangle $thus reduces to the recursion relation
\begin{eqnarray}
J\mathrm{e}^{-g}\Psi_{n+1}++ 2hq_n \Psi_n + J\mathrm{e}^{g} \Psi_{n-1}=E \Psi_n.
\end{eqnarray}
By introducing a transfer matrix, this Schr\"{o}dinger  equation can be written in a compact form by 
\begin{eqnarray}
\left(\begin{array}{c}
\Psi_{n+1} \\
\Psi_n
\end{array}\right)=M_{n}\left(\begin{array}{c}
\Psi_n \\
\Psi_{n-1}
\end{array}\right),
\end{eqnarray}
where the transfer matrix is given by
\begin{eqnarray}
M_{n}=\left(\begin{array}{cc}
J^{-1}\left(2hq_n-E\right)\mathrm{e}^{g} & -\mathrm{e}^{2g} \\
1 & 0
\end{array}\right).
\end{eqnarray}
According to the Oseledec's theorem, the exponent of  Lyapunov exponent $\gamma_n$ are the eigenvalues of a limiting matrix 
\begin{eqnarray}
\Gamma=\lim _{L \rightarrow \infty}\left(Q_L Q_L^{\dagger}\right)^{1 / 2 L},
\end{eqnarray}
where $Q_L=\prod_{n=1}^L M_n$. The smallest Lyapunov exponent describes the slowest growth of the wave function, and  also corresponds to the inverse of the localization length $\lambda$.  
In summary, the spectrum of the effective  Hamiltonian in the localized phase is a discrete spectrum of exponentially localized wave functions, whose characteristic localization length is given by the inverse of the Lyapunov exponent of the transfer matrix product.

\begin{center}
  {\centering \bf Supplementary Note 5: Non-Hermitian quantum disentangled liquid}
  \end{center}

  In the main text, we divide the system into three parts to make sure the two subsystems have same sizes for calculating the quantum mutual information. The area law scaling of fermions and volume law scaling of spins successfully  reveal the existence of non-Hermitian quantum disentangled liquid in the localized and delocalized phases. In order to exclude the effect of partitioning method on the non-Hermitian quantum disentangled liquid, we also provide the results where the sizes of subsystem are unequal $\mathrm{A} \neq \mathrm{B} $, see Supplementary Figure \ref{mutual-scaling} (a). Although the subsystems $A$ and  $B$ have the same numbers of fermion sites, comparing to the case where $A=B$, the  quantum mutual information of fermions at the nonequilibrium steady states are both increased in the localized phase (see Supplementary Figures \ref{mutual-scaling} (b)(c)) and delocalized phases (see Supplementary Figures \ref{mutual-scaling} (e)(f)), just because that the subsystem $A$ has one more bond spin. In the localized and delocalized phases, as shown in Supplementary Figures \ref{mutual-scaling} (d)(g), the area law scaling  of fermions and volume law scaling of spins are still clearly present.  The non-Hermitian quantum disentangled liquids are independent of the partitioning method of whole system.

\end{document}